\documentclass[fleqn,usenatbib,useAMS,twocolumn]{aastex63}
\usepackage[utf8]{inputenc}

\usepackage{color}
\usepackage{graphicx}
\usepackage{amsmath}
\usepackage{amssymb}
\usepackage{bm}
\usepackage{acronym}
\usepackage{ifthen}
\usepackage{blindtext}
\usepackage[normalem]{ulem}
\usepackage{hyperref}
\usepackage{etoolbox}
\usepackage{fancyhdr}
\usepackage{xspace}
\usepackage{textcomp}
\usepackage{multirow}
\usepackage[caption=false]{subfig}
\usepackage{lineno}
\usepackage{tabularx}
\usepackage{aas_macros} 

\def\gws{gravitational waves\xspace}

\newacro{S/N}{signal-to-noise ratio}

\newacro{PN}{post-Newtonian}

\newacro{O3}{the third observing run}

\newacro{O2}{the second observing run}

\newacro{MSP}{millisecond pulsar}

\newacro{CW}{\emph{Continuous Wave}}

\def\NICER{{\it NICER}\xspace}

\def\psrname{J0537\xspace}

\def\Hz{\textrm{Hz}}
\def\HzS{\textrm{Hz s$^{-1}$}}
\def\HzSsq{\textrm{Hz s$^{-2}$}}

\newcommand{\F}{{\mathcal F}}
\newcommand{\G}{{\mathcal G}}

\newcommand{\be}{\begin{equation}}
\newcommand{\ee}{\end{equation}}
\newcommand{\bea}{\begin{eqnarray}}
\newcommand{\eea}{\end{eqnarray}}

\newtoggle{fullauthorlist}
\toggletrue{fullauthorlist}

\shorttitle{\gws from r-modes in PSR~J0537\textminus6910}
\shortauthors{Abbott et al.}

\begin{document}
\title{Constraints from LIGO O3 data on gravitational-wave emission due to r-modes in the glitching pulsar PSR J0537\textminus6910}

\iftoggle{fullauthorlist}{
\AuthorCollaborationLimit=3000  


\author{R.~Abbott}
\affiliation{LIGO Laboratory, California Institute of Technology, Pasadena, CA 91125, USA}
\author{T.~D.~Abbott}
\affiliation{Louisiana State University, Baton Rouge, LA 70803, USA}
\author{S.~Abraham}
\affiliation{Inter-University Centre for Astronomy and Astrophysics, Pune 411007, India}
\author{F.~Acernese}
\affiliation{Dipartimento di Farmacia, Universit\`a di Salerno, I-84084 Fisciano, Salerno, Italy  }
\affiliation{INFN, Sezione di Napoli, Complesso Universitario di Monte S.Angelo, I-80126 Napoli, Italy  }
\author{K.~Ackley}
\affiliation{OzGrav, School of Physics \& Astronomy, Monash University, Clayton 3800, Victoria, Australia}
\author{A.~Adams}
\affiliation{Christopher Newport University, Newport News, VA 23606, USA}
\author{C.~Adams}
\affiliation{LIGO Livingston Observatory, Livingston, LA 70754, USA}
\author{R.~X.~Adhikari}
\affiliation{LIGO Laboratory, California Institute of Technology, Pasadena, CA 91125, USA}
\author{V.~B.~Adya}
\affiliation{OzGrav, Australian National University, Canberra, Australian Capital Territory 0200, Australia}
\author{C.~Affeldt}
\affiliation{Max Planck Institute for Gravitational Physics (Albert Einstein Institute), D-30167 Hannover, Germany}
\affiliation{Leibniz Universit\"at Hannover, D-30167 Hannover, Germany}
% ...
}

{
  \author{The LIGO Scientific Collaboration}
  \affiliation{LSC}
  \author{The Virgo Collaboration}
  \affiliation{Virgo}
  \author{The KAGRA Collaboration}
  \affiliation{KAGRA}
}
\date{\today}


\begin{abstract}
We present a search for continuous gravitational-wave emission due to r-modes in the pulsar PSR J0537\textminus6910 using data from the LIGO-Virgo Collaboration observing run O3. PSR~J0537\textminus6910 is a young energetic X-ray pulsar and is the most frequent glitcher known. The inter-glitch braking index of the pulsar suggests that gravitational-wave emission due to r-mode oscillations may play an important role in the spin evolution of this pulsar. Theoretical models confirm this possibility and predict emission at a level that can be probed by ground-based detectors. In order to explore this scenario, we search for r-mode emission in the epochs between glitches by using a contemporaneous timing ephemeris obtained from \textit{NICER} data. We do not detect any signals in the theoretically expected band of 86-97 Hz, and report upper limits on the amplitude of the gravitational waves. Our results improve on previous amplitude upper limits from r-modes in J0537-6910 by a factor of up to 3 and place stringent constraints on theoretical models for r-mode driven spin-down in PSR~J0537\textminus6910, especially for higher frequencies at which our results reach below the spin-down limit defined by energy conservation.
\end{abstract}

\keywords{gravitational waves --- pulsars: general --- pulsars: individual (PSR~J0537\textminus6910) --- stars: neutron}

\section{Introduction}

PSR~J0537\textminus6910 is a young (1-5 kyrs) energetic X-ray pulsar, rotating at a spin frequency $\nu=62$ Hz \citep{Marshall98}, in the Large Magellanic Cloud at a distance of 49.6 kpc \citep{Pietrzynski19}. PSR~J0537\textminus6910 (hereafter J0537) has been the subject of a number of studies, starting from its first detection with the \textit{Rossi X-ray Timing Explorer} (\textit {RXTE}; \citealt{Bradt93}) up to recent observations starting in 2017 with the \textit{Neutron Star Interior Composition Explorer} (\textit{NICER}; \citealt{Gendreau12}). 

J0537 is intriguing for several reasons. Not only is it the fastest spinning young pulsar known, but measurements of its spin evolution also reveal J0537 to be the most prolific glitcher known. The pulsar exhibits large glitches, i.e., sudden increases in spin frequency $\Delta\nu$ of approximate size $\Delta\nu/\nu\approx 10^{-7}$, roughly every 100 days \citep{Marshall04,Middleditch06,Antonopoulou18,Ferdman18,Ho20}. This in itself is already peculiar, as most pulsars do not glitch regularly. Rather, pulsars mostly have glitch size distributions that are consistent with power-laws and distributions for waiting times between glitches consistent with exponentials, with the Vela pulsar being one of the few other exceptions that glitches quasi-periodically \citep{Melatos08, Howitt18, Fuentes19}. J0537 is, however, unique, as it is the only glitching pulsar that shows a strong correlation between the size of a glitch and the waiting time to the next glitch \citep{Middleditch06, Antonopoulou18,Ferdman18,Ho20}, which suggests that a threshold has to be reached to trigger the glitch mechanism (see \citealt{HaskellMelatos15} for a review of pulsar glitch models).

One can try to understand the impact of glitches on the spin evolution of J0537 by comparing its long-term spin evolution, i.e., the trend over a number of years and consequently over many glitches, to its short-term spin evolution between glitches.
\citet{Antonopoulou18} studied the spin evolution over a 13-year span of \textit{RXTE} data (1999--2011) and determined a long-term second frequency derivative $\ddot{\nu}=-7.7\times 10^{-22}\mbox{ Hz s$^{-2}$}$ (and $\dot{\nu}\approx-1.99\times10^{-10}\mbox{ Hz s$^{-1}$}$), which leads to a braking index $n=\nu\ddot{\nu}/\dot{\nu}^2=-1.22\pm0.04$, while \cite{Ferdman18} measured $\ddot{\nu}=-8.2\times 10^{-22}\mbox{ Hz s$^{-2}$}$ and $n=-1.28\pm0.04$ with the same dataset.  More recently, \citet{Ho20} used 2.7 years of \textit{NICER} data to extend and refine these values to $\ddot{\nu}=-8.00\times 10^{-22}\mbox{ Hz s$^{-2}$}$ and $n=-1.25\pm0.01$.

The braking index $n$ is obtained by assuming a power-law spin-down mechanism for the neutron star of the form $\dot{\nu}\propto-\nu^n$, where $n=3$ if magnetic dipole radiation (at constant magnetic field strength and inclination) is the dominant spin-down mechanism. A negative value of $n$ thus describes an unusual spin evolution, which may be a consequence of the cumulative effect of glitches during the more than 20-year time span of monitoring observations since 1999 (see discussions in \citealt{Antonopoulou18, Ho20}). In order to test this hypothesis, it is  of interest to study the braking index between glitches. This allows us to understand if, far from a glitch, it is possible to extract an `intrinsic' braking index that can provide information on the physical spin-down mechanism for J0537. A detailed analysis of post-glitch relaxations shows that, while the inter-glitch braking index is large for days after a glitch, it tends to an asymptotic value of $n\approx 7.4$ for longer times \citep{Andersson18}. Similar values of $n$ are also obtained independently by \citet{Ferdman18} and from analysis of recent \textit{NICER} observations \citep{Ho20,Abbott21}. Such a value may be indicative of the spin evolution of J0537 not being driven by electromagnetic wave emission but by gravitational-wave (GW) emission due to a constant amplitude r-mode oscillation for which $n\approx 7$ \citep{Andersson18}.

Theoretically this is an intriguing possibility. The r-mode is a toroidal mode of fluid oscillation for which the restoring force is the Coriolis force (see \citealt{Haskell15} and \citealt{Kokkotas16} for a review), and it is generically unstable to GW emission, which can drive the mode to large amplitudes \citep{Andersson98, Friedman98}. In fact, it has been suggested that this is the mechanism that spins down young and hot neutron stars \citep{Lindblom98,Andersson99}: as their internal temperature drops, newborn neutron stars enter the region of parameter space in which viscosity cannot efficiently damp the r-mode and thus the mode grows to a large amplitude, and the ensuing GW emission drives the spin-down of the system. \citet{Alford14} showed that this evolution ends quite generically for any hadronic equation of state at spin periods in the range observed for the standard pulsar population. Furthermore \citet{Alford14} showed that, of the known systems, J0537 is the only one that could theoretically still be in the last phases of its r-mode driven evolution. This has also been confirmed by \citet{Andersson18}, who find that indeed an unstable r-mode could be present in the star and drive its spin-down, although the amplitude, depending on the unknown mass and equation of state of the star, would have to be close to the upper end of theoretical estimates for the saturation amplitude. 

At such oscillation amplitudes, r-mode driven GW emission is an interesting source for ground-based interferometers, which could potentially confirm or reject the scenario discussed above. A first search for r-mode emission from J0537 was carried out by \citet{Fesik20,fesikpapa20b} using data from the O1 and O2 runs of the Advanced LIGO network; they did not detect a signal but obtained upper limits on the amplitude of the mode within an order of magnitude of the spin-down limit (i.e., the value that would be needed to explain the entire observed spin-down of the star by GW energy loss). Additionally, a search for GW emission at both once and twice the rotational frequency (i.e., for emission due to `mountains') of J0537 was carried out using LIGO and Virgo O2 and O3 data \citep{Abbott21}, which did not detect a signal but probed for the first time below the spin-down limit for this alternative GW emission mechanism.
A search for GWs from r-modes in the Crab pulsar has recently been reported by 
\citet{rajbhandari2021searches}, with a setup similar to the one used here. 
No GWs were detected, but upper limits beating the spin-down limit were obtained.

In this paper we present results of searches for r-mode GW signals from J0537 in O3 data, where we make use of timing information obtained from {\it NICER} observations of J0537 that cover the O3 period \citep{Ho20}. The search was conducted using two independent approaches, the time domain  
$\F$/$\G$ - statistic method \citep{1998PhRvD..58f3001J,Jaranowski:2010} and the 5-vector method \citep{nb_method, nb_method_plus}.

\section{Data sets and ephemeris}
In this section we briefly describe both the GW data that have been used 
in the searches and the ephemeris of pulsar J0537 that is needed for the analysis.

\subsection{Gravitational-wave data}\label{sec:gwdata}
\begin{figure}[htb]
    \centering
    \includegraphics[width=\columnwidth]{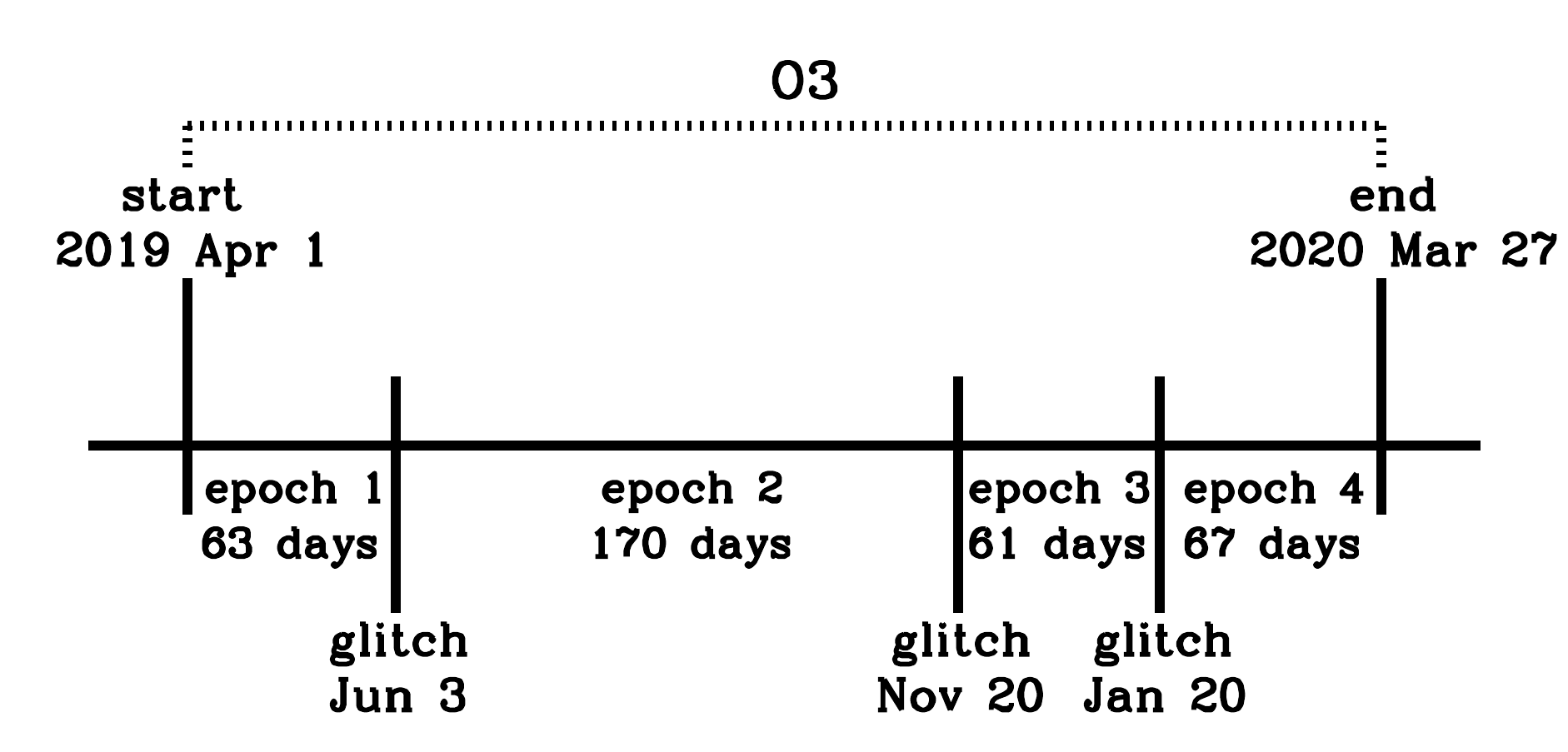}
    \caption{Timeline of O3 observing run, glitches, and epochs between glitches.
    } 
    \label{fig:timeline}
\end{figure}

We use data from the third observing run (O3) of the two Advanced LIGO detectors \citep{2015CQGra..32g4001L}. The O3 data run lasted from 2019 April 1 to 2020 March 27, with a one-month pause in data collection in October 2019 (see Figure~\ref{fig:timeline}). The Hanford (H1) and the Livingston (L1) detectors had duty factors of $\sim 76\%$ and $\sim 77\%$, respectively during O3.
For the LIGO O3 data set, the analysis uses the calibration with estimated upper limits on amplitude and phase uncertainties of $\sim 7\%$ and $\sim 4 \deg$, respectively \citep{Sun_2020}, which we use as conservative estimates of the true calibration uncertainty near the frequencies analyzed here.

\subsection{PSR~J0537\textminus6910 ephemeris}\label{sec:nicer}

The timing analysis of J0537 using \NICER data from 2017 August 17 to 2020 April 25 is described in \citet{Ho20}. During this timespan eight glitches were detected, with the last three glitches occurring during the O3 run (see Figure~\ref{fig:timeline}).  These three glitches divide up the O3 dataset into four inter-glitch epochs, which form natural segments into which to divide the data analyses in the searches described below.  Note that parameters of a revised ephemeris for epoch 4, which are presented in \citet{Abbott21}, are within the errors of the ephemeris from \citet{Ho20} and thus do not affect our analysis here. In the first of our two search methods to be presented ($\F$/$\G$ statistic search), the data is indeed divided into four segments in this way.  In the second search described (5-vector method), the long $\sim 170$~day inter-glitch epoch is divided into two for computational reasons, giving a total of five data segments in the analysis.  More detail will be given below.

\section{Gravitational-wave radiation from r-modes}\label{sec:signal}

\subsection{Gravitational-wave frequency} \label{sec:gwf}

The r-mode GW emission frequency $f$ depends on the pulsar spin frequency $\nu$ and on the neutron star structure (e.g., \citealt{Idrisy15}).  Following the analysis of \citet{carideetal19}, the relationship can be expressed as
\begin{eqnarray}
\label{eq:f_val}
f&=& A \nu - B \left( \frac{\nu^2} {\nu_K^2} \right) \nu,
\\
\label{eq:f_dot_val}
\dot{f}&=& A \dot{\nu} - 3B \left( \frac{\nu^2} {\nu_K^2} \right) \dot{\nu},
\\
\label{eq:f2}
\ddot f&=&A \ddot\nu - \left( 3 + \frac{6}{n} \right) B \frac{\nu^2} {\nu_K^2}  \ddot\nu .
\end{eqnarray}
where the dots denote time derivatives of the frequencies and $\nu_K$ is the Keplerian frequency for which centrifugal forces would tear the star apart. These relations make allowance for fully relativistic gravity (via the $A$-parameter) and for rotation of the star (via the $B$-parameter) and are thus likely to be accurate as long as no other physical mechanism plays a significant role in the r-mode oscillation, e.g., a resonance between the core and crust oscillation frequencies (see \citealt{Levin01}).  The precise way in which these relations are used in this paper differs slightly between the two searches we present, as will be described below.

The values of $A$ and $B$ for any particular neutron star depend both upon its (unknown) mass, and also upon the (unknown)  Equation of State (EoS) of dense matter.  \citet{carideetal19} considered a range of astrophysically motivated neutron star masses,  and a range of microphysically motivated equations of state, to suggest a conservative range of values of $A$ and $B$ to employ in a search:
\begin{eqnarray}
\label{eq:A_range}
1.39 < &A& < 1.57 , \\
\label{eq:B_range}
0 < &B& < 0.195 .
\end{eqnarray}

One also needs a value of the (unknown) Keplerian frequency $\nu_K$ to make use of the above equations.  We again follow \citet{carideetal19} in making a conservative choice, taking the (low) value $\nu_K\simeq 506$ Hz. This value is obtained by assuming that the Keplerian frequency coincides with the frequency of the most rapidly rotating neutron star currently known, 716 Hz for PSR J1748-2446ad \citep{Hessels06}, and that J0537 is a low mass star, with a consequently lower $\nu_K$; see \citet{carideetal19} for relevant discussion.

\subsection{Signal model} \label{sect:signal_model}

The GW signal from an isolated  rotating neutron star is described by the model derived in \citet{1998PhRvD..58f3001J}. The signal depends on four extrinsic parameters and  $3 + s$ intrinsic parameters, where $s$ is the number of spin-down parameters included in the model.
The four extrinsic parameters are the GW amplitude $h_0$, the star's spin axis inclination angle (with respect to the line of sight) $\iota$, the GW polarization angle $\psi$, and a constant phase $\phi_0$. 
The $3 + s$ intrinsic parameters are the frequency $f_0$,
$s$ spin-down parameters $f_1,\dots,f_s$, and two angles describing the position in the sky of the source, i.e., right ascension $\alpha$ and declination $\delta$.
For GW emission from r-modes, the only difference from the model of \citet{1998PhRvD..58f3001J} is that the polarization angle $\psi$ needs to be transformed as
\begin{equation}\label{eq:psi}
\psi \rightarrow \psi + \pi/4.
\end{equation}
The derivation of this transformation is given and discussed  in detail in \citet{Owen:2010}.
The position in the sky of J0537 is precisely known (J2000; \citealt{Townsley06}):
\be\label{eq:Jpos}
\alpha = 05^{\mathrm h}37^{\mathrm m}47.\!^\mathrm{s}416, \  \delta = -69^\circ  10\arcmin 19\farcs88.
\ee

The intrinsic parameters determine the phase modulation $\phi(t)$ of the signal, which in the model of \citet{1998PhRvD..58f3001J} is approximated by
\be\label{eq:phase}
\phi(t) \cong 2\pi\sum_{k=0}^{s}f_k \frac{t^{k+1}}{(k+1)!}
+ \frac{2\pi}{c} \mathbf{n}(\delta,\alpha)\cdot\mathbf{r}_\text{d}(t)
\sum_{k=0}^{s}f_k\frac{t^k}{k!},
\ee
where $\mathbf{r}_\text{d}(t)$ is the position of the detector in the solar system barycenter frame and $\mathbf{n}$ is the constant unit vector in the direction of the neutron star in that frame and is
\be\label{eq:n_eq}
\mathbf{n} = 
(\cos\delta\cos\alpha,\cos\delta\sin\alpha,\sin\delta).
\ee
This formula takes into account both the intrinsic spin-down of the source and the Doppler modulations produced by the motion of the detector.

In addition to providing information on timing, electromagnetic observations also provide insight into the likely orientation of the spin axis of the pulsar. J0537 is associated with the supernova remnant N157B in the Large Magellanic Cloud. Observations of the pulsar wind nebulae (PWN) torus of the remnant have enabled an accurate measurement of the orientation  of the axis of symmetry of the torus \citep{Ng:2008}. It is very likely that
the symmetry axis of the torus coincides with the spin axis of the pulsar. This enables us to estimate the values of the two parameters of the GW signal: polarization angle $\psi$ and inclination angle
$\iota$. For the former, we have  $\psi = 2.2864\pm0.0384$~rad. From the observations, we cannot obtain the absolute direction of the rotation of the pulsar: we have two possible values of the inclination angle $\iota$ or $\pi - \iota$ \citep{2010MNRAS.402.2503J}. Thus the measurements yield $\iota = 1.522\pm0.016$~rad or $\iota = 1.620\pm0.016$~rad.  This information can be used to carry out versions of our searches where this prior information on these two angles is incorporated into the search, as will be described below. 
An inclination angle $\iota$ close to $\pi/2$ means that the GW signal is almost linearly polarized and consequently it has nearly the smallest maximum signal-to-noise ratio achievable. 

\section{Search methods}

Two different search methods are used: the \emph{time domain $\F$/$\G$-statistic} method and the \emph{5-vector} method.  
In the following, we describe their main features and the slightly different parameter space they cover.

\subsection{Time domain $\F$/$\G$-statistic method}

{\em Detection statistics.} We perform a coherent search of time domain segments of the data spanning between the glitches of the pulsar. The coherent search is performed using the $\F$-statistic derived in \citet{1998PhRvD..58f3001J}. The $\F$-statistic has been used extensively in searches for GWs from rotating neutron stars, including the recent directed searches for GWs from r-mode emission by  \citet{Fesik20,fesikpapa20b}.
In this search, we use an efficient implementation of the $\F$-statistic presented in \citet{Astone:2010,Aasi:2014}. Its two main features are the use of  the fast Fourier transform (FFT) to evaluate the $\F$-statistic over the whole band of the data analysed and the use of a grid of templates based on optimal coverings of the  parameter space. The FFT computation is enabled by a technique called resampling which consists of interpolating the data to Solar System Barycenter time (see Section IIID of \citet{1998PhRvD..58f3001J} for details). The resampling needs to be performed for all frequency derivatives only once for each sky position that we analyze. Thus in the directed search presented here, where the pulsar's position is known to a very high accuracy, the resampling needs to be performed only once.

If the angles $\psi$ and $\iota$ are known, there is a variant of the $\F$-statistic called the $\G$-statistic \citep{Jaranowski:2010} which implements the coherent search in this case.  As described in Section  \ref{sect:signal_model} above, for \psrname the inclination angle $\iota$ is very close to $\pi/2$, which  indicates that the GW signal is almost linearly polarized.
We perform a Monte Carlo simulation where, for an observation time of 60 days, we add simulated signals to Gaussian noise
with the two possible values of the inclination angle $\iota$ of \psrname consistent with observations of the PWN. We then search for the signal in both cases with a template for $\iota = 1.522$. We perform the simulation for an array of signal-to-noise ratios (SNRs) from 5 to 25. We find no appreciable difference in the detectability of the signals and accuracy of the estimation of the frequency parameters for the two cases. The SNR of the signals recovered using the template with a ``wrong'' inclination angle decreased by no more than 2\% and the  accuracy of the estimations of frequency and spin-down rate decreased by no more than 2\%. Thus we conclude that only one value of the inclination angle is needed when using the $\G$-statistic method for the search.

In this paper, we analyze the data using both the $\F$ and $\G$-statistic. We search the data from the two LIGO detectors and consequently use a multi-detector implementation of the statistics developed in \citet{krolak:2004}, \citet{krolak:2007}, \citet{JK:2009}, and \citet{Cutler:2005hc}.

\begin{figure}[htb]
    \centering
    \includegraphics[width=\columnwidth]{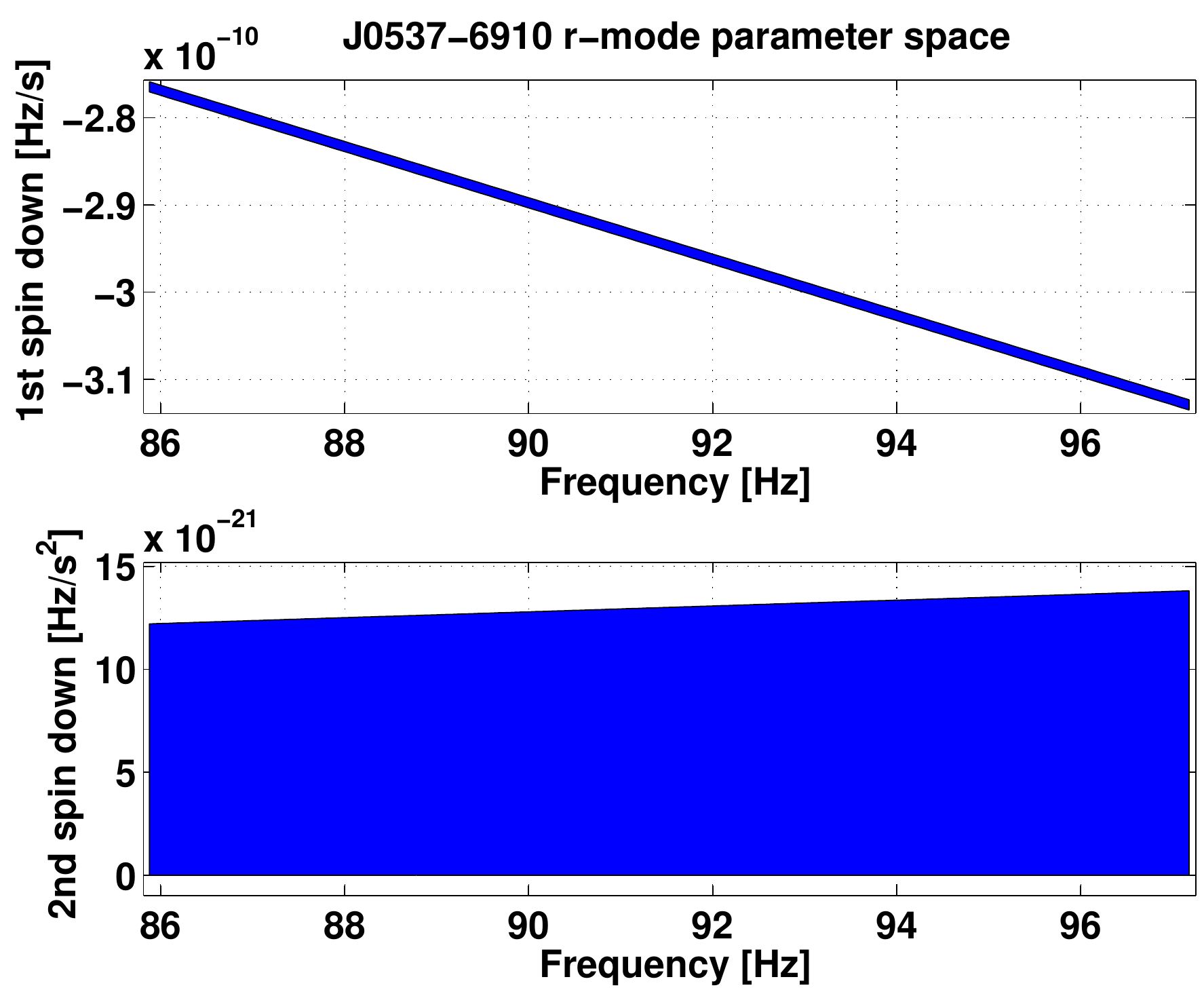}
    \caption{Parameter space $f-\dot{f}$ (top panel) and $f-\ddot{f}$ (bottom panel) for r-mode GW emission from pulsar J0537
determined by Eqs. (\ref{range0}), (\ref{range1}), and (\ref{range2}).  We assume  values of the spin parameters $\nu$, $\dot\nu$, and $\ddot\nu$ of the pulsar from Table 1 of \citet{Ho20} for the epoch at MJD 58723.} 
    \label{fig:ParSpace}
\end{figure}

{\em Parameter space}. The relation between stellar spin frequency $\nu$ and r-mode GW frequency $f$ is described in equations (\ref{eq:f_val}--\ref{eq:f2}), following the model of  \citet{carideetal19}, who used these relations to propose ranges of GW frequencies and frequency derivatives over which to search: 
\begin{eqnarray}
\label{range0}
\nu \left( A_{\min} - B_{\max} \frac{\nu^2} {\nu_K^2} \right)
\leq&f&\leq \nu\, A_{\max},
\\
\label{range1}
-\dot\nu \left( \frac{f}{\nu} - 2B_{\max} \frac{\nu^2}
{\nu_K^2} \right) \leq& - \dot f&\leq -\dot\nu \frac{f}{\nu},
\\
\label{range2}
0 \leq&\ddot f&\leq \ddot\nu \frac{f}{\nu},
\end{eqnarray}
where $\ddot\nu$ is the maximum value consistent with electromagnetic observations.
The value used for $\nu_K$ and the maximum and minimum values we assume for $A$ and $B$ are given in Section \ref{sec:gwf}.
The parameter space for the frequency and its two derivatives defined by the above equations is illustrated in Figure \ref{fig:ParSpace}.

In the J0537 r-mode search  of  \citet{Fesik20,fesikpapa20b} the prescription for the parameter space proposed by \citet{carideetal19}
was also used, however their assumed ranges of  $\dot f$  and  $\ddot f$  were independent of  the frequency $f$.

{\em Grid of templates.} The problem of constructing a grid in the parameter space is equivalent to the {\em covering problem} \citep{conway-99,Prix:2007}. There exist optimal coverings that have the smallest possible number of grid points per unit volume of the parameter space. Our implementation of the $\F$ and $\G$-statistic uses an FFT which evaluates the statistic at Fourier frequencies which in general do not coincide with nodes of the optimal grid. To address this problem, we introduce constrained grids compatible with application of the FFT \citep{Astone:2010}. For the case of the directed search that we present here, in which the position of the source in the sky is known, the construction of the constrained grids is described in \citet{pisarski:2011}. 

One important problem is to determine how many spin-down parameters $f_k$ 
(see Eq.~\ref{eq:phase}) we need to take into account in the search. This problem is addressed in \citet{brady-98}. In general, a longer  observation time $T$ requires the inclusion of more spin-down parameters in our templates so that we can match the signal accurately. Thus we calculate the number of grid points $N_k(T)$ for a given number $k$ of spin-down parameters included as a function of observation time $T$.
The observation time at which we need to include an additional $f_{k+1}$ spin-down parameter is the observation time at which the curves $N_k(T)$  and $N_{k+1}(T)$ intersect (see Figure 6
of \citet{brady-98} for illustration).

\begin{figure}[htb]
    \centering
    \includegraphics[width=\columnwidth]{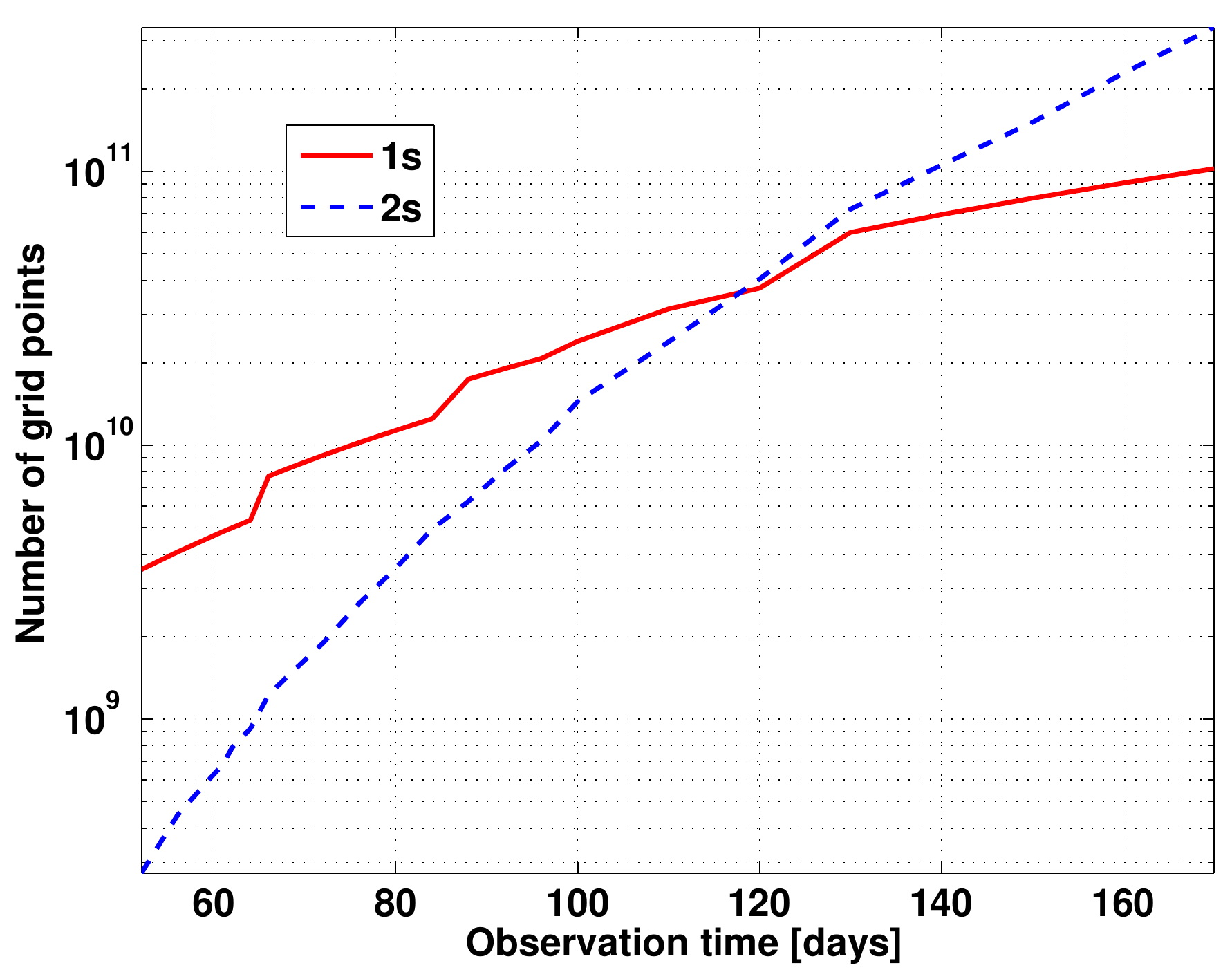}
    \caption{Number of templates as a function of the observation time for the case when one spin-down parameter $f_1$ (continuous line) and two spin-down parameters $f_1$ and $f_2$ (dashed line) are included.}
    \label{fig:grid}
\end{figure}

The results for this search are shown in Figure \ref{fig:grid}. 
We see that the number of grid points $N_2(T)$ when $f_2$ is included becomes larger than the number of grid points $N_1(T)$ when only $f_1$ is included for observation times $T$ greater than 117 days. In this analysis we assume a minimum spin-down age of 5000 years, appropriate for the  pulsar J0537. Thus for the expected range of frequency parameters of the GW signal from r-modes from \psrname, we need to include $f_2$ in the templates only for coherent observation times longer than 117 days. 

To support this result, we perform a Monte Carlo simulation for an observation time of 66 days where we compare the detectability of the GW signal with one spin-down and two spin-down parameters using the $\F$-statistic with templates which include only one spin-down parameter. The simulation is performed for Gaussian noise and for an array of SNRs from 5 to 25.
We find no appreciable difference in the detectability of the signals and in the accuracy of estimation of the frequency and the first spin-down parameter, whether or not the second  spin-down parameter is included in the simulated signals.  The SNR of the signals recovered when two spin-down parameters are included decreased by no more than 1\% with respect to signals when only one spin-down parameter is included. The accuracy of the estimation of the frequency and the spin-down parameter decreased by no more than 1\%.

{\em Threshold and false alarm probability.} In order to assess the significance of the candidates obtained during the search, we need to calculate the false alarm probability. We follow the approach developed in \citet{JK:2009} (see Chapter 6.1.3). In that approach, the parameter space that we search is divided into a number $N_c$ of cells. The cells are defined in such a way that the autocovariance function of the statistic between a point inside a given cell and outside that cell is less than $\frac{1}{2}$.
This means that the statistic between cells is approximately uncorrelated and the cells can be considered as independent. Thus the probability $P^T_F$ that a statistic 
${\mathcal L}$ exceeds a threshold ${\mathcal L}_o$ in \textit{one or more} cells is given by
\be\label{eq:FP}
P^T_F({\mathcal L}_o) = 1 - \big[1 - P_F({\mathcal L}_o)\big]^{N_c},
\ee
where $P_F({\mathcal L}_o)$ is probability distribution of the statistic at a given point 
in the parameter space assuming that data is only noise. This is by definition 
the false alarm probability (FAP). The number of cells $N_c$ is defined as
\be\label{eq:NC}
N_c = \frac{V}{v_c},
\ee
where $V$ is the volume of the parameter space and $v_c$ is the volume of the cell. 
By inverting Eq.~(\ref{eq:FP}), we can obtain the threshold ${\mathcal L}_o$ for a chosen value of the false alarm probability.
To calculate the thresholds using Eq.~(\ref{eq:FP}), we assume Gaussian noise, i.e., we assume that the $2 \F$ and $2 \G$-statistics have central $\chi^2$ distribution with 4 and 2  degrees of freedom, respectively. \newline

{\em The search setup.}  During the O3 run, three glitches of \psrname are detected, as described in Section \ref{sec:nicer}. These are glitches 6--8 presented in Tables 1 and 2 of 
\citet{Ho20} (see also \citealt{Abbott21}). For our analysis, we  select four
data segments between the times of the three glitches of the pulsar, with the first segment starting at the beginning of the O3 run and the last segment
ending at the end of O3. We select data segments that have length equal to an integer multiple of sidereal days. The start times of the segments after the glitches that occurred in O3 are chosen 
to be 1 day after the glitch and the end times around 1 day before the glitch. This results in data segments 62, 168, 59 and 66 sidereal days long. 

\begin{deluxetable*}{ccccc}[htb]
\tablecaption{Parameters of the search of the four time domain segments analyzed using the $\F$ / $\G$-statistic method.
\label{tab:SearchSetUp}}
\tablewidth{0pt}
\tablehead{
\colhead{Segment} & \colhead{1} & \colhead{2} & \colhead{3} & \colhead{4}
}
\startdata
$t_0$ [GPS] & 1238163456 & 1243641549 & 1258329549 & 1263599949 \\
$T_{\rm span}$ [days] & 62 & 168 & 59 & 66 \\
$\F_o$ & 12.5 & 15  & 12.5 & 12.5 \\
$\G_o$ & 10   & 12.5 & 10 & 10 \\
$\delta f$ [Hz] & $1.19 \times 10^{-7}$ & $5.96 \times 10^{-8}$ & $1.19 \times 10^{-7}$  & $1.19 \times 10^{-7}$  \\
$\delta \dot{f}$ [Hz s$^{-1}$] & $2.21 \times 10^{-14}$ & $1.30 \times 10^{-14}$ & $2.28 \times 10^{-14}$ & $1.43 \times 10^{-14}$ \\
$\delta \ddot{f}$ [Hz s$^{-2}$] & - & $3.14 \times 10^{-21}$ & - & - \\
$N_t$ & $5.02 \times 10^9$ & $6.09 \times 10^{10}$ & $4.54 \times 10^9$ & $7.74 \times 10^9$  \\
$\F_{0.01}$ & 26.2 & 26.3 & 26.5 & 26.9 \\
$\G_{0.01}$ & 22.9 & 23.0 & 23.2 & 23.6 \\
\enddata
\end{deluxetable*}

The initial GPS times $t_0$ of each data segment are given in Table~\ref{tab:SearchSetUp}. During the longest 168-day segment we were observing for only 138 days because of the commissioning break in October 2019.
In each time domain segment, we search for GW emission due to r-modes in the parameter space determined by Eqs.~(\ref{range0}), (\ref{range1}), and (\ref{range2}). We adopt values of the spin parameters $\nu$, $\dot\nu$, and $\ddot\nu$ of J0537 from Table 1 of \citet{Ho20}.
For segments 62, 168, 59, and 66 days long, we select spin
parameters for epochs at MJD~58600, 58723, 58836, and 58918, respectively. 
We include the second spin-down derivative only for 168 day segments because, as shown above, for the other segments the second spin-down parameter can be ignored.
The search is performed using both the $\F$- and $\G$-statistic.
For the $\G$-statistic, we fix angles $\psi$ and $\iota$ at
\be
\psi = 2.2864, \ \iota = 1.522.
\ee
In Table \ref{tab:SearchSetUp}, we give thresholds $\F_o$ and $\G_o$ used for the $\F$- and $\G$-statistic search, respectively. We also present grid spacings $\delta f$, $\delta \dot{f}$, and $\delta \ddot{f}$
in frequency and two spin-down parameters as well as the number $N_t$ of  templates used. Moreover in Table \ref{tab:SearchSetUp}, we show thresholds $\F_{0.01}$ and $\G_{0.01}$ for the $\F$- and $\G$-statistic, respectively, corresponding to a false alarm probability of 1\% for searches in each individual segment. These thresholds are used to select candidates from the search for follow-up analysis.

\subsection{5-vector method}
\label{sec:fivemeth}
The {\it 5-vector} method has its foundation in the coherent pipeline for the {\it narrow-band} search of continuous waves (CWs) from known pulsars \citep{nb_method, nb_method_plus}, which has been applied to the analysis of both LIGO and Virgo data \citep{nb_vsr4, nb_O1, nb_O2}. {\it Narrow-band} searches are motivated by the need to take into account a possible mismatch between the actual GW frequency and the value inferred from EM observations. The mismatch could arise from errors in the EM measurements or in the presence of differential rotation if the GW signal is mainly emitted by the inner parts of the star, rather than by the crust, or if free precession is excited. Since the EoS is uncertain (see Section~\ref{sec:signal}), so is the GW frequency. As a consequence, we need to search over a wide range of frequencies (and corresponding spin-down rates).

The {\it 5-vector} r-mode search is based on an extension of the standard {\it narrow-band} pipeline. The main reason why the standard pipeline is not directly applicable for this search is the presence of glitches. The {\it 5-vector} pipeline is a fully coherent algorithm: it assumes that the phase evolution is regular and therefore a Taylor expansion involving the frequency and its derivatives can be defined \citep{nb_method_plus}. In the presence of glitches, this hypothesis is only valid within each intra-glitch period. In fact, it is not possible to perform a coherent analysis along the whole GW observing run, since there are phase discontinuities at the times of each glitch. To overcome this problem, a semi-coherent search is performed: intra-glitch periods are studied coherently and independently, and the results are later combined incoherently. This is only possible thanks to {\it NICER} observations, which provided the glitch times.

Other relevant modifications are implemented in order to deal with a parameter space larger by a factor $\sim 10^4$ with respect to standard {\it narrow-band} searches. This is due to an enlargement factor of $\sim$ 200 for the frequency and $\sim$ 50 for spin-down rate. Thus, optimization schemes to manage the computational load have been developed.

{\em Parameter Space.} The equations relating r-mode GW frequency to spin frequency are given by equations~(\ref{eq:f_val})--(\ref{eq:f2}), following the model of  \citet{carideetal19}.  
The ranges in $f$ and $\dot f$ are obtained by inverting equation~(\ref{eq:f_val}) and (\ref{eq:f_dot_val}) in terms of A and B:
\begin{eqnarray}
\label{eq:A_val}
A&=&\frac{1}{2} \left( 3 \frac{f} {\nu} - \frac{\dot f} {\dot \nu} \right),
\\
\label{eq:B_val}
B&=&\frac{\nu^2_K}{2 \nu^2} \left(  \frac{f} {\nu} - \frac{\dot f} {\dot \nu} \right).
\end{eqnarray}
The region to investigate in the parameter space of $f$ and $\dot f$ is defined by the corresponding values of $A$ and $B$ that satisfy the relations $1.39 \leq A \leq 1.57$ and $0 \leq B \leq 0.195$, as described in Section \ref{sec:gwf}. These equations produce a parallelogram shape in parameter space, as shown in Figure~\ref{fig:5v_cand_par_space}, and reduce the volume to be searched over by more than a factor of 30 with respect to the choice of a rectangular grid in $(f,~\dot f)$. The $(f,~\dot f)$ region covered by this search is slightly different, at the corners, from that of the $\F / \G$ method, as can be seen comparing Figures \ref{fig:ParSpace} and \ref{fig:5v_cand_par_space}. The central step of the 5-vector pipeline is the application of spin-down rate corrections. Then, to cover this region in the parameter space a different frequency band is selected for each value of the spin-down rate. The correct frequency range for each spin-down rate is obtained by solving equation~(\ref{eq:f_dot_val}) for $A$ and substituting it into equation~(\ref{eq:f_val}) in order to get
\begin{equation}
    \label{eq_sd_freq_range}
    f=  \frac{\dot f}{\dot \nu} \nu + 2B \left( \frac{\nu^2} {\nu_K^2} \right) \nu.
\end{equation}
Using the assumed ranges of $A$ and $B$, we search the band 86-97~Hz, with a frequency step of $\delta f=1/T$, while
the spin-down is discretized in bins of width $\delta \dot{f}=1/T^2$.
Note that the second derivative $\ddot f$ can be considered as constant (within one frequency bin) for periods of length $T<T_{\rm max}$, with $T_{\rm max} = (2/ \ddot f_{\rm max})^{1/3}$, where $\ddot f_{\rm max}=\ddot \nu (A_{\rm max} - B_{\rm min} [ \frac{\nu^2} {\nu_K^2})]$ is the maximum value allowed for the second derivative of the GW frequency. Assuming a braking index $\approx7$, we can write $\ddot \nu = 7 \dot \nu^2/\nu $ which gives $T_{\rm max} \approx 75$ days. These considerations lead to the choice of the data segments to be studied coherently and independently, as reported in Table~\ref{tab:IGPeriods}. Segments 2 and 3 actually correspond to the same intra-glitch period (same rotational parameters), which we split in order not to exceed the maximum length $T_{\rm max}$. Therefore, the second derivative is fixed for all segments for the 5-vectors pipeline to the value $\ddot f = \bar A \ddot \nu= 1.48\times10^{-20}\mbox{ Hz s$^{-2}$}$, where $\bar A$ is the average value for the first-order parameter $A$ (second-order corrections are neglected to fix this value). We fix $\ddot \nu= 1\times10^{-20}\mbox{ Hz s$^{-2}$}$ for all segments, which is consistent with {\it NICER} measurements reported in \citet{Ho20}. The second-order spin-down bin is $\delta \ddot{f}\simeq 2/T^3=1.30\times10^{-20}\mbox{ Hz s$^{-2}$}$ and is large enough to include the uncertainty in $\ddot{f}$ due to the unknown exact values of parameters $A$ and $B$ and to the measurement of $\ddot{\nu}$. 
The main search parameters are shown in Table~\ref{tab:IGPeriods}.

\begin{deluxetable*}{ccccccccc}[htb]
\tablecaption{Parameters of the five time domain periods analyzed using the 5-vectors pipeline. We have $\delta \dot f=\delta f ^2$. The chosen time periods are the same over which \textit{NICER} provides the pulsar ephemeris. 
\label{tab:IGPeriods}}
\tablewidth{0pt}
\tablehead{
\colhead{Period} & \colhead{MJD Start} & \colhead{MJD End} & \colhead{MJD Epoch} & \colhead{Duration} & \colhead{$\nu$ [Hz]}  & \colhead{$\dot \nu$ [$10^{-10}$ Hz s$^{-1}$]} & \colhead{Number of points} & \colhead{$\delta f$ [$10^{-7}$ Hz]}  
}
\startdata
1 & 58574 & 58629 & 58600 & 55 & 61.9145 & $-1.9974$ & $1.54 \times 10^9$ & 2.1898 \\
2 & 58646 & 58701 & 58723 & 55 & 61.9124 & $-1.9973$ & $1.63 \times 10^9$ & 2.1492 \\
3 & 58701 & 58757 & 58723 & 56 & 61.9124 & $-1.9973$ & $1.63 \times 10^9$ & 2.1492 \\
4 & 58810 & 58863 & 58836 & 53 & 61.9104 & $-1.9974$ & $1.29 \times 10^9$ & 2.3212 \\
5 & 58873 & 58935 & 58918 & 62 & 61.9090 & $-1.9977$ & $2.35 \times 10^9$ & 1.9026 \\
\enddata
\end{deluxetable*}

{\em Search pipeline description.}  The full search band of 86--97~Hz is divided into 1-Hz intervals for computational reasons. For each interval,
the coherent part of the search pipeline, which consists of different steps \citep{nb_method, nb_method_plus}, is applied. Schematically:  application of the Doppler and Einstein delay corrections\footnote{This is done by means of {\it resampling}, so that a single correction holds for the whole frequency band.}; application of spin-down corrections for each bin of the spin-down grid via a {\it heterodyning} signal processing technique; computation of network detection statistic using 5-vectors. These are complex arrays containing the Fourier Transform of the data\footnote{Actually the Fourier Transform is computed by means of Fast Fourier Transform (FFT) algorithm.} at the five frequencies at which the power of a CW signal, emitted at a given frequency, would be split by sidereal modulation. 
In order to reduce the effect of spectral leakage due to frequency discretization, which can produce a sensitivity loss of up to $\sim 36\%$, we use an `interbinning' procedure \citep{nb_method_plus}. This  consists of estimating the data FFT values at the halfway bin $k+\frac{1}{2}$ in terms of the values at the $k$-th and $(k+1)$-th bins as
\begin{equation}
    \tilde{x}_{FFT,k+\frac{1}{2}}\approx \frac{\pi}{4}\left(\tilde{x}_k - \tilde{x}_{k+1}\right).
    \label{eq:interbin}
\end{equation}
The detection statistic is computed separately on the natural grid and on the grid of shifted bins.

The incoherent part consists of tracking the evolution of the GW signal through the glitches. For each point of the parameter space of the first segment with GW parameters $(f, \dot f)$, we first compute the corresponding $(A, B)$ using equations~(\ref{eq:A_val}) and (\ref{eq:B_val}). Then keeping $A$ and $B$ constant, the GW frequency/spin-down rate pair $(f, \dot{f})$ in the other segments is constructed by taking into account the change of rotational parameters due to glitches. In practice, a different grid in the plane $(f, \dot{f})$ needs to be used for each segment, as the duration of each segment is in general different (see Table~\ref{tab:IGPeriods}). When a pair $(f, \dot f)$ is extrapolated from one segment to the following segment, the pair corresponding to the point of the grid closest to the extrapolated value is taken. At the end, for each point in  the parameter space of the first segment, a weighted average of the corresponding detection statistic over the five analyzed segments is computed. The weights are computed by the inverse of the median of the data power spectrum in the 1~Hz band to which the point belongs. 
The fundamental assumption we make in this procedure is that the coefficients $A$ and $B$ do not change across the glitches. As their value is related to the neutron star equation of state and thus should not be affected by glitches, this assumption is reasonable. \\
Once the overall weighted statistic is computed for each point of the parameter space, the loudest outliers (i.e., those having the highest value of the detection statistic) are selected every 1~mHz over the corresponding whole spin-down range. For each outlier, the significance is computed through the p-value, obtained by comparing the outlier statistic value to the statistic noise distribution, following the procedure used in \citet{nb_O1} and \citet{nb_O2}. More specifically, the noise distribution is computed from the data itself, taking the 1~Hz band which includes the outlier, with the exclusion of the 1~mHz band to which the outlier belongs, plus the preceding and the following 1~Hz bands. Outliers with p-value smaller than 1$\%$ (after properly taking into account the trials factor) are upgraded to the level of candidates and followed up in order to increase their significance or, conversely, to demonstrate they are incompatible with an astrophysical GW signal.

\section{Results}

\subsection{$\F$ / $\G$ - statistic results} 
\begin{figure}[htb]
    \centering
    \includegraphics[width=\columnwidth]{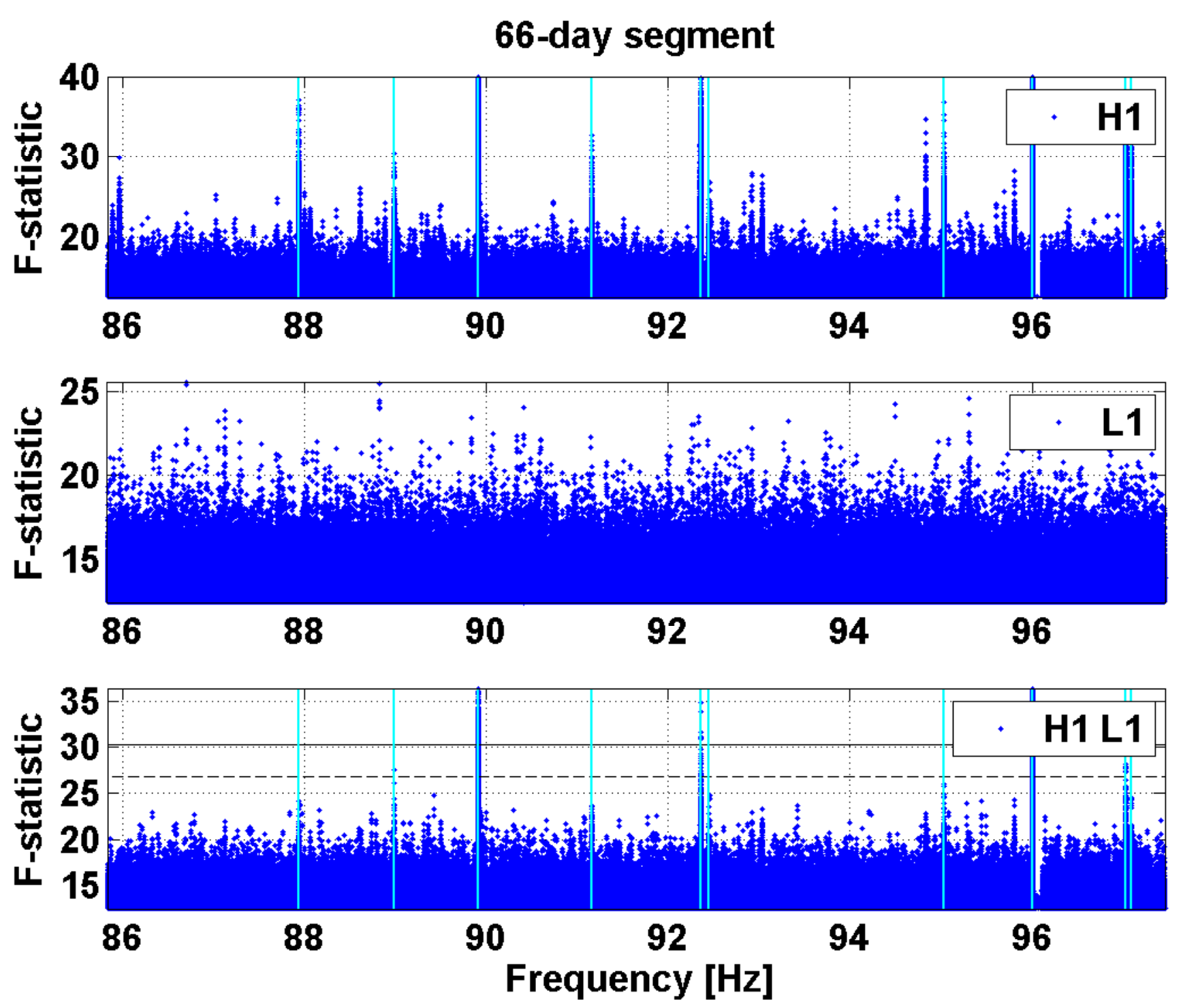}
    \caption{Candidates from the search of the 66-day segment using the $\F$-statistic. The top panel are candidates from Hanford detector (H1) data search, the middle panel are candidates from Livingston detector (L1) data search, and the bottom panel from the network search (H1 L1). We see that there are several lines present in the Hanford data that contaminate the network search. The lines are marked by vertical lines. The continuous horizontal line corresponds to $0.1\%$ false alarm probability for the whole O3 parameter space, and the dashed horizontal line is for $1\%$ false alarm probability for the search in the 66-day segment.}
    \label{fig:F66all}
\end{figure}

{\em Candidates.} We performed the search in LIGO data for GW emission from
r-modes in \psrname during the four time-domain segments using both the $\F$ and $\G$-statistic methods. In the frequency band searched, the data contain several known periodic interferences that correlate well with our templates. To identify the origin of the interferences in each case, we also searched the data from the Hanford and Livingston detectors separately. In Figure~\ref{fig:F66all}, we present results of the search for the case of 66-day segments using the $\F$-statistic.
We see in this case several strong lines are present in the Hanford detector data. These lines are not present in Livingston detector data but contaminate the search of the network of detectors. Most of these lines are integer multiples of 1~Hz. As these single-detector lines cannot be of GW origin, we veto them before further analysis of the candidates. The lines vetoed in the case of the 66-day segment are marked by vertical lines
in Figure~\ref{fig:F66all}.
In searches of all data segments, the strong lines that led to threshold-crossing outliers only appeared in the Hanford detector data.

\begin{figure}[htb]
    \centering
    \includegraphics[width=\columnwidth]{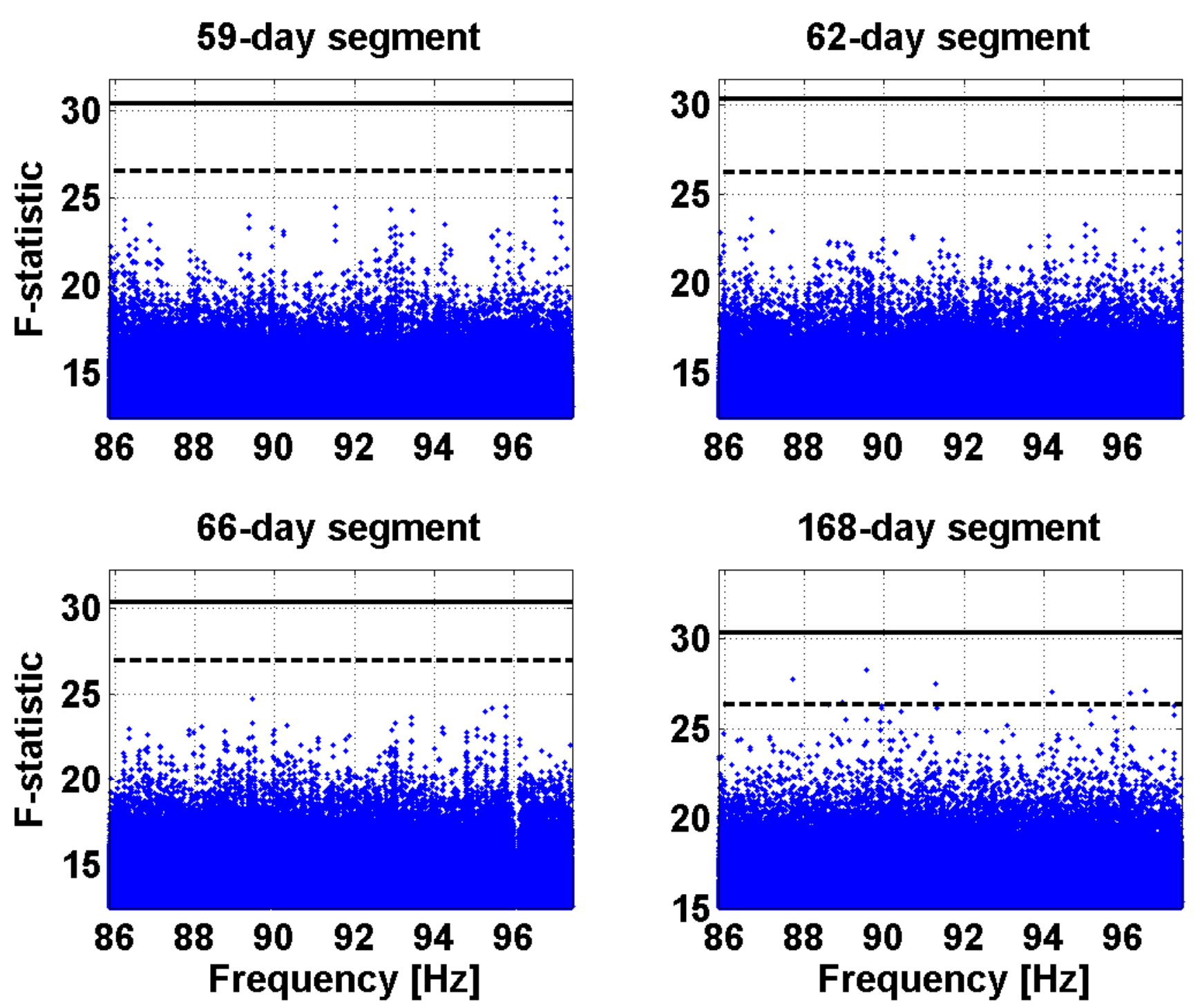}
    \caption{Candidates from the search of the four segments using the $\F$-statistic. The continuous horizontal lines give threshold for the $\F$-statistic corresponding to 
    $0.1\%$ false alarm probability for the search in the whole O3 parameter space. 
    The dashed horizontal lines 
    are drawn at threshold $\F_{0.01}$ of the $\F$-statistic corresponding 
    to $1\%$ false alarm probability for the search in each segment. 
    The values of $\F_{0.01}$ for each segment are given in Table~\ref{tab:SearchSetUp}.}
    \label{fig:F62}
\end{figure}

\begin{figure}[htb]
    \centering
    \includegraphics[width=\columnwidth]{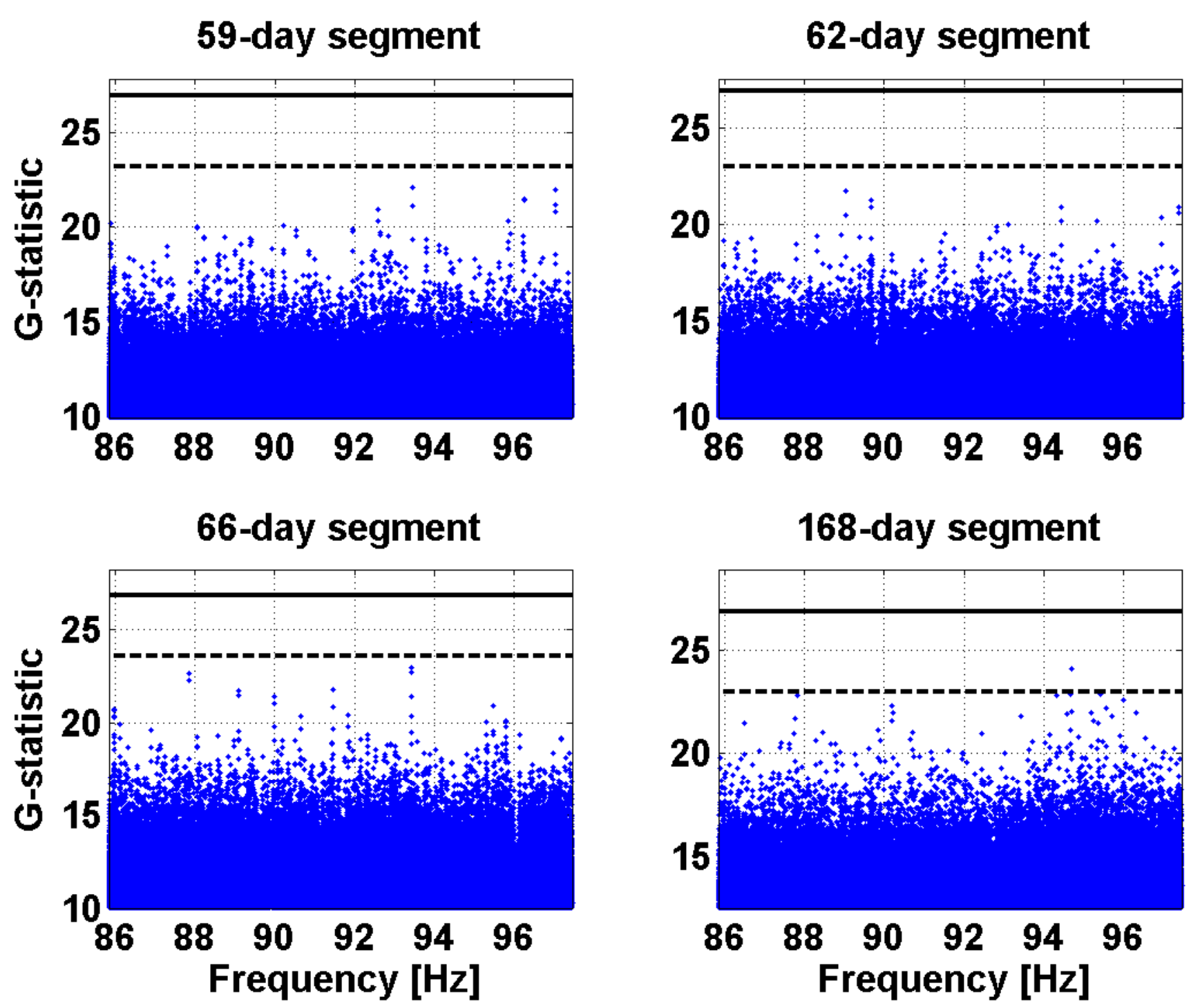}
    \caption{Candidates from the search of the four segments using the $\G$-statistic. The continuous horizontal lines give threshold for the $\G$-statistic corresponding to 
    $0.1\%$ false alarm probability for the search in the whole O3 parameter space. 
    The dashed horizontal lines 
    are drawn at threshold $\G_{0.01}$ of the $\G$-statistic corresponding 
    to $1\%$ false alarm probability for the search in each segment. 
    The values of $\G_{0.01}$ for each segment are given in Table~\ref{tab:SearchSetUp}.}
    \label{fig:G62}
\end{figure}

In Figure~\ref{fig:F62} we present candidates obtained in our search of the four time-domain segments using the $\F$-statistic and in Figure~\ref{fig:G62} candidates using the $\G$-statistic. 
These candidates are after vetoing the lines. The continuous horizontal lines correspond to 
the threshold for $0.1\%$ false alarm probability.
The threshold is calculated for the total number of cells in all the four time-domain segments. 

We also give thresholds corresponding to $1\%$ false alarm probability for the parameter space searched limited to each individual segment. These thresholds are drawn as dashed horizontal lines. 

We see that none of the candidates is strong enough to cross the global threshold for the false alarm probability 
of $0.1\%$.  However there are several crossings of the $1\%$ local threshold for the 168-day segment. These subthreshold candidates are listed in Table~\ref{tab:SubCandidates}, together with values of their frequency and spin-down parameters. The SNRs of the candidates are also given.

\begin{deluxetable}{cccccc}[htb]
\tablecaption{\label{tab:SubCandidates} Subthreshold candidates for the $\F$/$\G$ - statistic search.}  
\tablehead{ 
   \colhead{$f$ [$\Hz$]} & \colhead{$\dot{f}$ [$\HzS$]} & \colhead{$\ddot{f}$ [$\HzSsq$]} & \colhead{SNR}} 
\startdata
\multicolumn{4}{c}{$\G$-statistic 168 days} \\ \hline  
94.7057 & $-3.04 \times 10^{-10}$ & $1.04 \times 10^{-20}$ & 6.8 \\
\multicolumn{4}{c}{$\F$-statistic 168 days} \\ \hline  
87.6870 & $-2.82 \times 10^{-10}$  & $3.47 \times 10^{-21}$ & 7.2 \\
89.5434 & $-2.88 \times 10^{-10}$  & $6.94 \times 10^{-21}$ & 7.2 \\
88.9333 & $-2.86 \times 10^{-10}$  & $3.47 \times 10^{-21}$ & 7.0 \\
88.5916 & $-2.85 \times 10^{-10}$  & $3.47 \times 10^{-21}$ & 7.0 \\
91.3000 & $-2.94 \times 10^{-10}$  & $6.94 \times 10^{-21}$ & 7.1 \\
94.2063 & $-3.04 \times 10^{-10}$  & $3.47 \times 10^{-21}$ & 7.1 \\
96.5544 & $-3.11 \times 10^{-10}$  &      0  & 7.1 \\
96.1719 & $-3.09 \times 10^{-10}$  & $3.47 \times 10^{-21}$ & 7.1 \\
\enddata  
\end{deluxetable}  

Subthreshold candidates occur only in the 168-day segment.
We searched for coincident candidates to the subthreshold candidates in the three other data segments, correcting the frequency and spin-down parameters of the candidates for glitches and frequency evolution.
No coincident candidate is found, and therefore we conclude that we have no valid candidate for a GW signal in our search. 

\begin{figure}[htb]
\centering
    \includegraphics[width=\columnwidth]{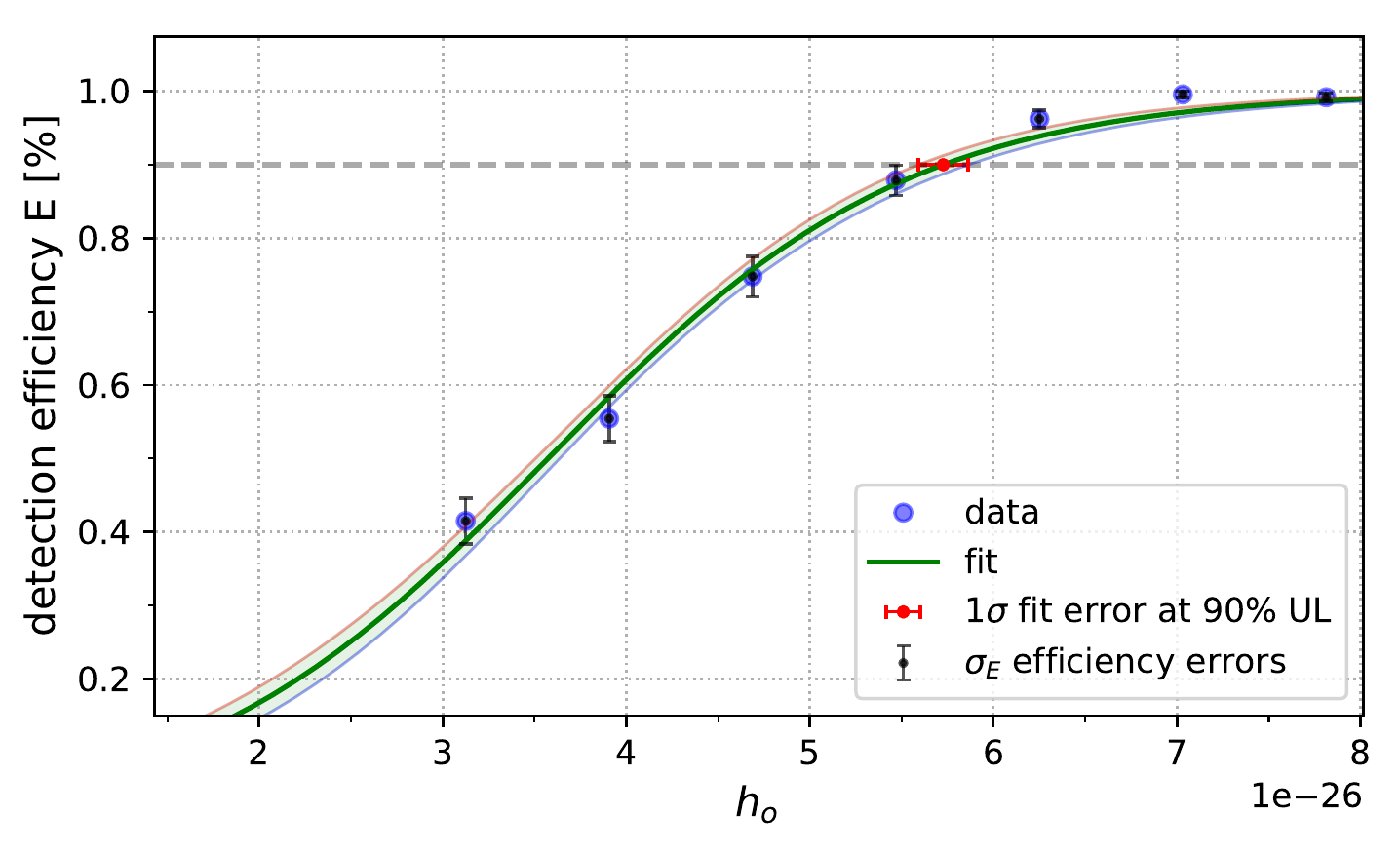}
    \caption{Example sigmoid function fit (green solid line) to the injected data efficiencies (blue dots), representing the detection efficiency $E$ as a function of injected GW amplitude $h_o$. Pale red and blue curves mark the $1\sigma$ confidence band obtained from the uncertainty of the fit. Red error bar marks the ${\pm}1\sigma$ standard deviation on the $h_o^{90\%}$ value, corresponding to the efficiency of 0.9 (indicated by the horizontal dashed gray line). Vertical errors for each efficiency represent $1\sigma$ standard binomial errors related to detection rate, $\sigma_E = \sqrt{E(1-E)/N_i}$, where $E$ is the efficiency and $N_i=250$ is the number of injections for each GW amplitude. The data is evaluated with the $\F$-statistic method in the 59~day segment at the mid-band frequency of 91.423~Hz.}
    \label{fig:ul_sigmoid_example}
\end{figure}

{\em Upper limits and sensitivity of the search.}
As no significant candidate signal for GW emission from r-modes in \psrname is found, we set upper limits on the intrinsic GW amplitude $h_0$. We set upper limits for 0.5~Hz bands. In each band, we generate signals for an array of amplitudes $h_0$ for the position of J0537. For each amplitude, we generate 250 signals with $f$, $\dot{f}$, and $\ddot{f}$ (for the 168-day segments) chosen from uniform random distributions in their respective ranges. Moreover for the case of the $\F$-statistic search, we also choose the values of the polarization angle $\psi$ and cosine of the inclination angle $\iota$ from uniform random distributions. The signals are added to the real data and searches are performed with the same grids and search set-up as for the real data search in the neighbourhood of simulated signal parameters. We search $\sim \pm 3$ grid points for $\dot{f}$ and $\sim\pm 1$ grid points for $\ddot{f}$ from the parameters of the simulated signal. A signal is considered as detected if the highest detection statistic value from the simulated signal search is higher than the highest detection statistic obtained in the search of real data for a given 0.5~Hz band. The detection efficiency is the fraction of recovered signals. We estimate the $h_0^{90\%}$, i.e., 90\% confidence upper limit on the GW amplitude $h_0$, by fitting\footnote{We use the {\tt python 3} \citep{10.5555/1593511} {\tt scipy-optimize} \citep{2020SciPy-NMeth} {\tt curve\_fit} package, implementing the Levenberg-Marquardt least squares algorithm, to obtain the best fitted parameters, $x_0$ and $k$, to the sigmoid function. Errors of parameters $\delta x_0$ and $\delta k$ are obtained from the covariance matrix and used to calculate the standard deviation $\sigma_e$ of the detection efficiency as a function of $h_0$ i.e., the confidence bands around the central values of the fit. In practice, we use the {\tt uncertainties} package \citep{uncertainties} to obtain the ${\pm}1\sigma$ standard deviation on the $h_0$ value.} a sigmoid function to a range of detection efficiencies $E$ as a function of injected amplitudes $h_0$, $E(h_0) = \left(1 + \exp(k(x_0 - h_0))\right)^{-1}$, with $k$ and $x_0$ being the parameters of the fit. Figure~\ref{fig:ul_sigmoid_example} presents an example fit to the simulated data, with the $1\sigma$ errors on the $h_0^{90\%}$ estimate marked in red, for the evaluation of the $\F$-statistic method in the 59~day segment at the mid-band frequency of 91.423~Hz.

\begin{figure*}[htb]
\centering
    \includegraphics[width=2.1\columnwidth]{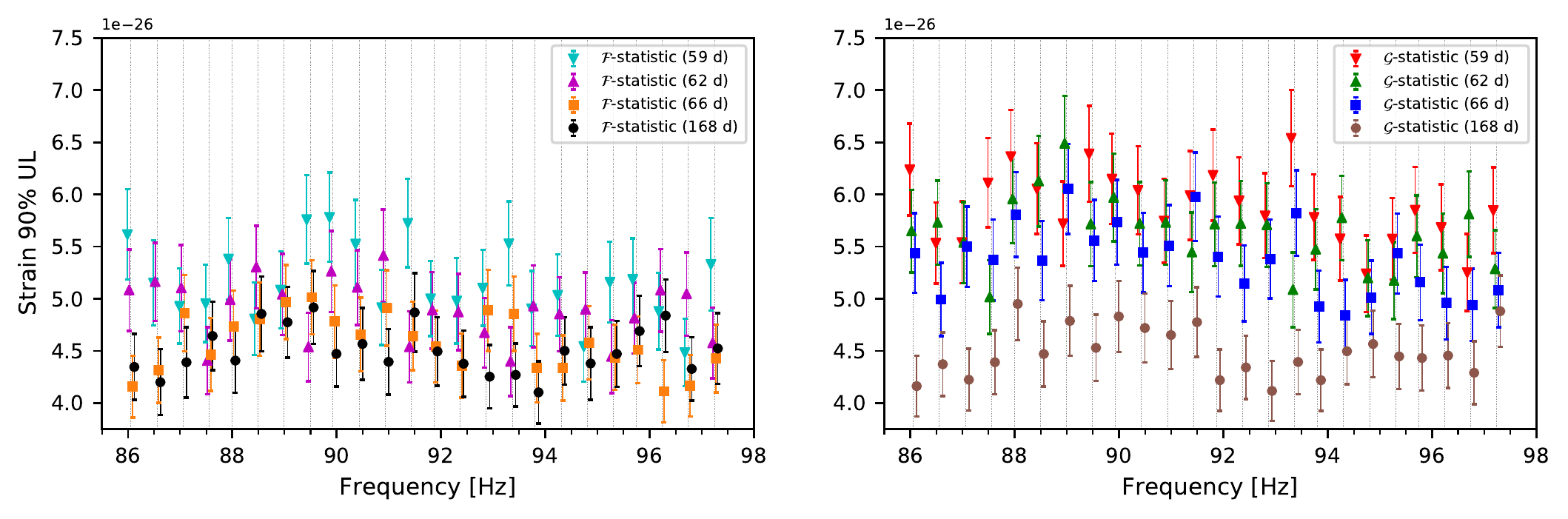}
    \caption{GW strain 90\% upper limits as a function of the band frequency for the $\F$-statistic method (left panel) and the $\G$-statistic method (right panel) with $1\sigma$ errors combining in quadrature the Monte Carlo simulation errors in amplitude ($1\sigma$ errors of the sigmoid fit) with the detectors' calibration uncertainty (see Sect.~\ref{sec:gwdata}). Thin vertical lines mark the mid-band frequencies. Points are shifted from their mid-frequency values to improve visibility.}
    \label{fig:lvk_f_vs_gstat}
\end{figure*}

These upper limits mean that a GW signal with amplitude $h_0^{90\%}$ from r-mode emission in \psrname  would be detected in our search with $90\%$ probability. Specifically, errors due to the Monte Carlo simulations with $N_i=250$ injections per amplitude are less than 3\% for the $\F$-statistic and less than 1\% for the $\G$-statistic results. In Fig.~\ref{fig:lvk_f_vs_gstat}, we present our $90\%$ upper limits for all four searches and for both $\F$- and $\G$-statistics, with the Monte Carlo simulation errors added in quadrature to the amplitude calibration uncertainty error of 7\% (see Sect.~\ref{sec:gwdata}); the latter dominates the uncertainty of our errors estimate. 
The upper limits for the $\G$-statistic are greater than for the $\F$-statistic. This is because for the $\G$-statistic the inclination angle $\iota$ is close to $\pi/2$ meaning that the GW signal is almost linearly polarized and consequently it has nearly the smallest maximum signal-to-noise ratio achievable what degrades the detection sensitivity. 

\begin{figure*}[ht]
\centering
    \includegraphics[width=2.1\columnwidth]{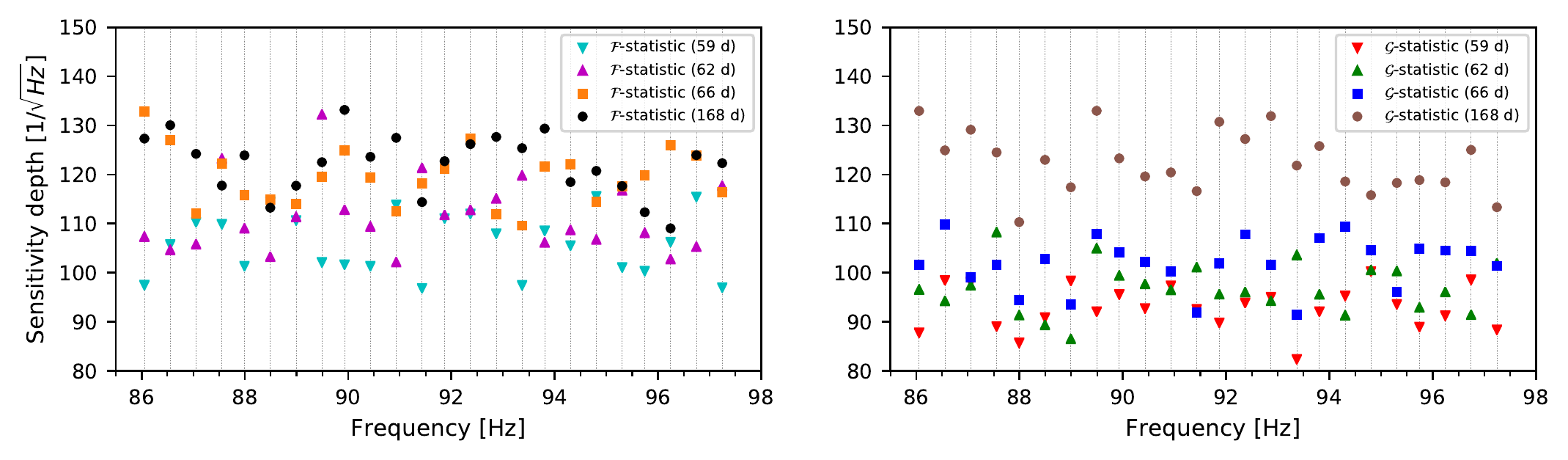}
    \caption{Sensitivity depth for searches of the four time domain segments using the $\F$-statistic (left panel) and the $\G$-statistic (right panel). Thin vertical lines mark the mid-band frequencies.}
    \label{fig:sens_depth_fgstat}
\end{figure*}

A useful measure of the performance of a search is the {\em sensitivity depth}, first introduced in \citet{2015PhRvD..91f4007B} and defined as
\begin{equation}
{{\mathcal{D}}}^{90\%}(f)={\frac{\sqrt{S_h(f)}}{h_o^{90\%}} }~~[ {1/\sqrt{\text{Hz}}} ],
\label{eq:sensDepth}
\end{equation}
where $\sqrt{S_h(f)}$ is the noise level associated with the signal frequency $f$.  
For a network of detectors, we take $S_h(f)$ as the harmonic mean of the single-detector one-sided power spectral densities of the data. Spectral densities of each detector are averaged over the 0.5~Hz frequency bands that the upper limits value refer to. The resulting ${{\mathcal{D}}}^{90\%}(f)$ values for the $\F$-statistic and $\G$-statistic analysis are shown in Fig.~\ref{fig:sens_depth_fgstat}.

\subsection{5-vector method results}

\begin{figure}[htb]
    \centering
    \includegraphics[width=\columnwidth]{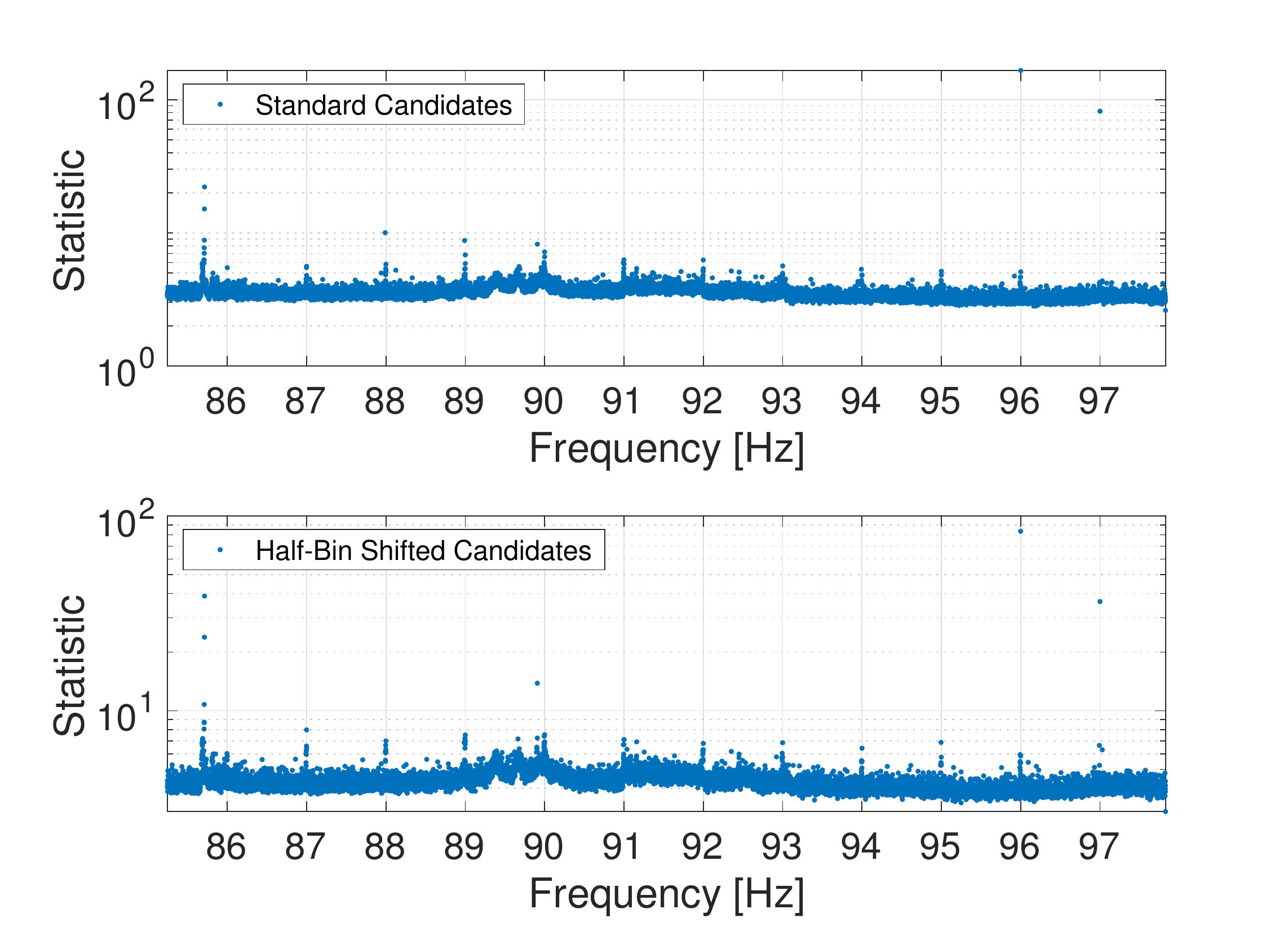}
    \caption{Normalized detection statistic for the full set of outliers selected by the 5-vector method on the natural and on the shifted search grid. Excess value of the detection statistic, due to detector disturbances, are evident at several integer frequencies.}
    \label{fig:5v_cand}
\end{figure}

{\em Outliers.} Outliers are obtained by selecting the loudest point in each millihertz of the studied frequency range. In fact, following the discussion in Section~\ref{sec:fivemeth}, two sets of outliers are selected: those with frequency in the natural grid and those with frequency belonging to the grid of shifted bins. The two sets of outliers are shown in the top and bottom panels of Figure \ref{fig:5v_cand}, respectively. 

\begin{figure}[ht]
    \centering
    \includegraphics[width=\columnwidth]{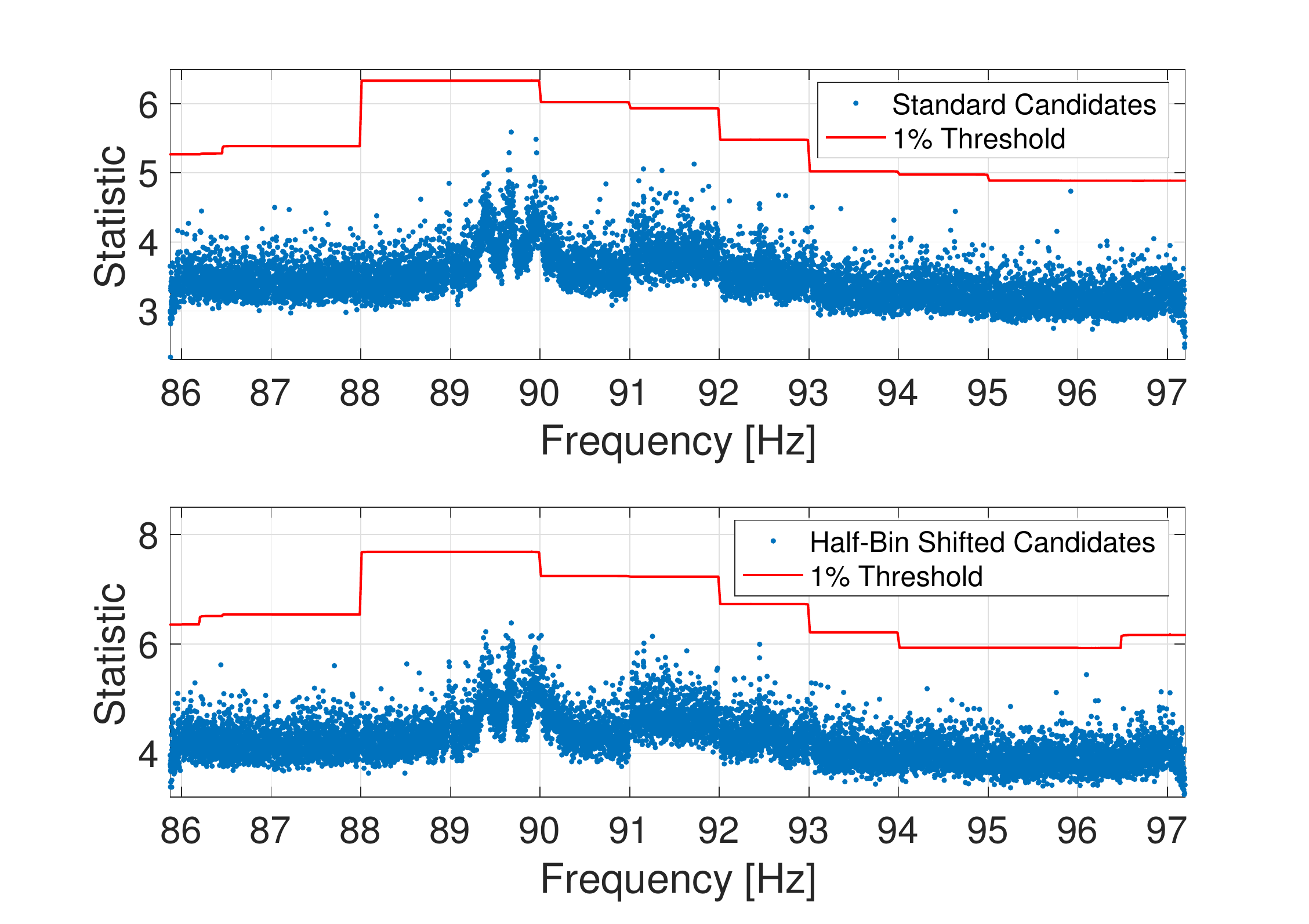}
    \caption{Normalized detection statistic for outliers selected in the natural frequency grid by the 5-vectors method, after the application of the vetoing criteria discussed in the text. The top panel refers to the natural frequency grid, while the bottom panel refers to the shifted frequency grid. The red continuous line represents the 1$\%$ p-value threshold on the detection statistic used to identify significant outliers.}
    \label{fig:5v_cand_vetos}
\end{figure}

It is clear from the figures that high values of the detection statistic correspond to several integer frequencies, as found also by the $\F/\G$ statistic method. They are due to disturbances of instrumental origin which mainly affect the Hanford detector\footnote{A list of known instrumental spectral disturbances can be found at https://www.gw-openscience.org/.}. 
In addition, to ensure computational accuracy, the search was initially performed slightly enlarging the physical ranges of the parameters (i.e., $1.38 \leq A \leq 1.58$ and $-0.01 \leq B \leq 0.205$).
To obtain more significant results, three sets of outliers are excluded:
 \begin{enumerate}
     \item outliers closer than 0.01~Hz to integer frequencies, in order to remove the impact of these disturbances on the noise distribution.
     \item outliers outside the physical ranges $1.39 \leq A \leq 1.57$ and $0 \leq B \leq 0.195$.
     \item outliers whose detection statistic in one segment is greater than the sum of the statistics on all the other four segments\footnote{We expect that, even if the r-mode is re-activated after each occurring glitch, it is very unlikely that the excitation mechanism changes to the point that the gravitational energy dissipated in a single period might be higher than the total energy emitted in all the other periods. On the other hand, it is more plausible to have temporary noise disturbances that might occur for a limited period of time, thus affecting only one of the studied periods.}. This veto limits the effect of temporary noise disturbances.
 \end{enumerate}
The outliers that remain after imposing these selection criteria are shown in Figure \ref{fig:5v_cand_vetos} for the natural frequency grid and for the shifted one. 
Outlier significance is estimated by means of the p-value. First, an overall p-value threshold, $p_{\rm thr}$, is computed such that $p_{\rm thr}\times N_p=0.01$, where $N_p$ is the total number of points in the explored parameter space. Then, for each outlier, the detection statistic noise distribution has been built considering the corresponding 1 Hz band, after the exclusion of the 1~mHz sub-band to which the outlier belongs, plus the preceding and following 1~Hz bands. On this distribution the threshold on the detection statistic corresponding to $p_{\rm thr}$ has been computed and used to establish if that outlier is highly significant. The detection statistic threshold is shown in Figure \ref{fig:5v_cand_vetos} as a continuous red line and is not surpassed by any of the outliers. As before, two sets of outliers are plotted: those selected in the natural frequency grid (top panel) and those selected in the shifted grid (bottom panel).

The detection statistic values distribution across the explored frequency range is not uniform due to the presence of noise artefacts (see e.g., the triple peak in the range 89-90 Hz) and to the globally non-flat detector noise curve, which decreases as the frequency increases.  

\begin{figure}[htb]
    \centering
    \includegraphics[width=\columnwidth]{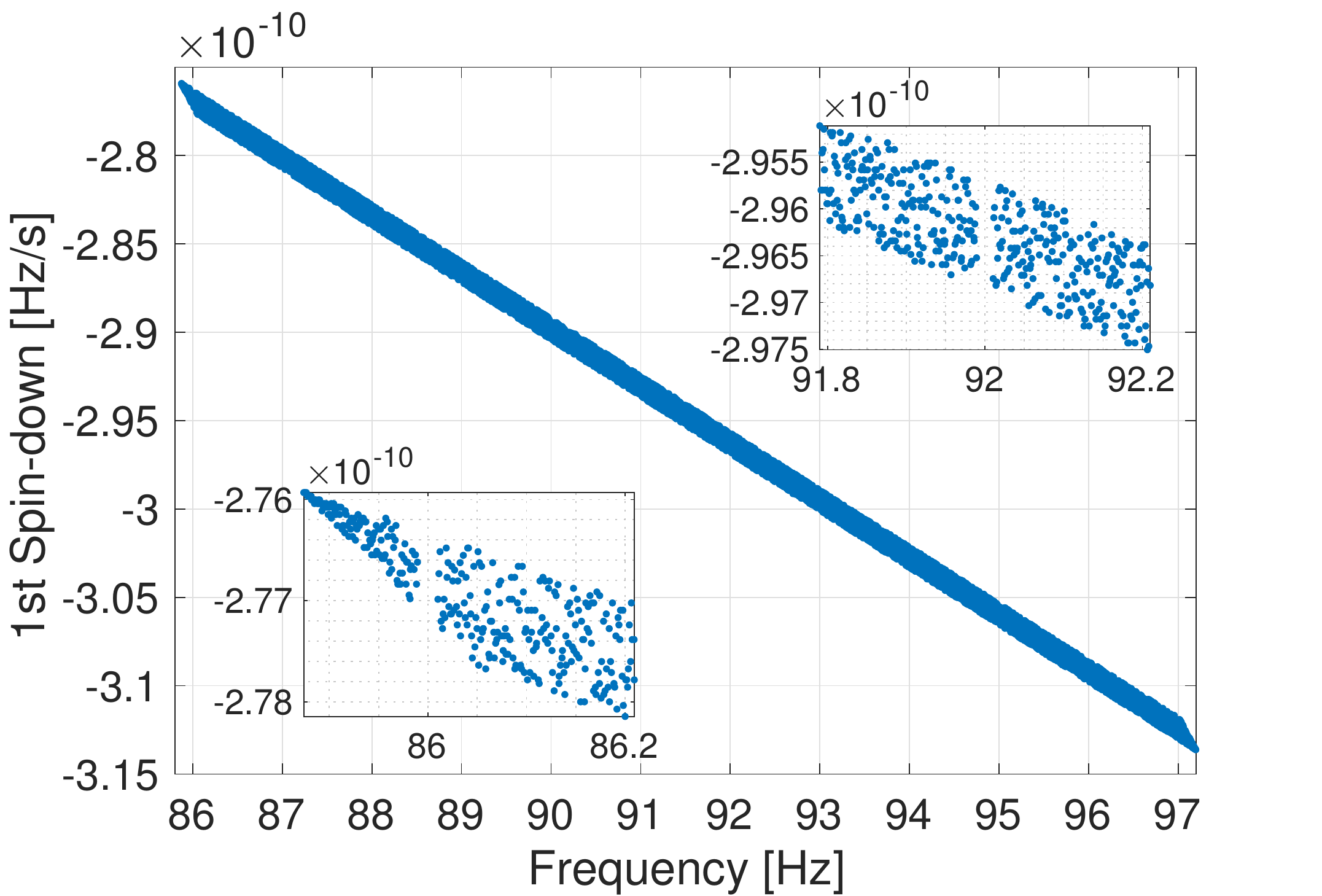}
    \caption{Outliers in the frequency/spin-down space selected by the 5-vectors method after vetoing criteria have been applied. In the two insets, we highlight the 0.02 Hz gaps around the integer frequencies (due to the presence of detector disturbances). It is also possible to see the shape of the regions at the edges of the frequency bands, that is slightly different from the one studied using the $\F$ / $\G$ statistic pipeline (reported in Figure \ref{fig:ParSpace}). The shape of these regions is constrained from Equations \ref{eq:A_val} and \ref{eq:B_val} using the ranges given by Equations \ref{eq:A_range} and \ref{eq:B_range}.}
    \label{fig:5v_cand_par_space}
\end{figure}

\begin{figure}[htb]
    \centering
    \includegraphics[width=\columnwidth]{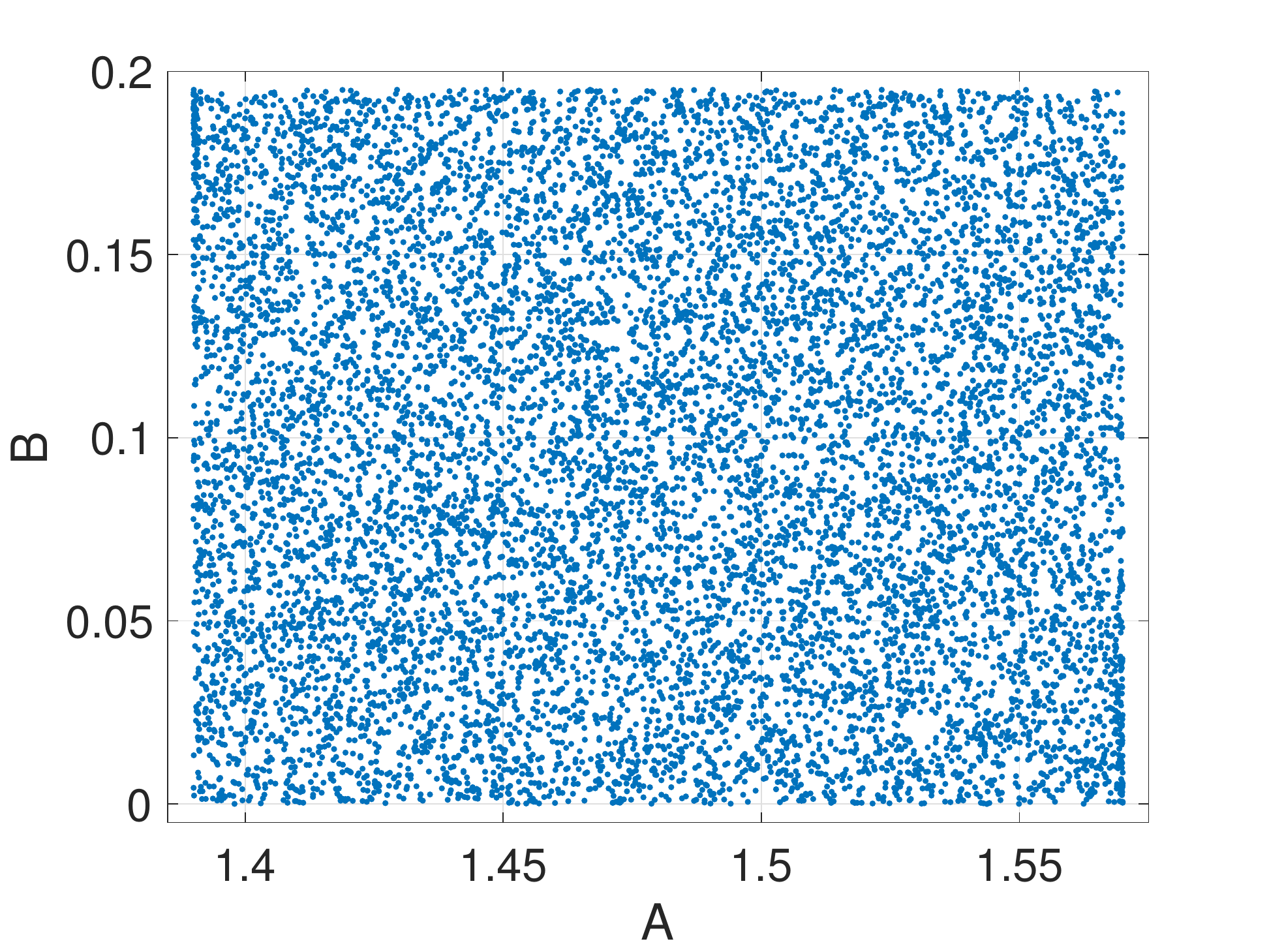}
    \caption{Outliers in the space of physical parameters $(A, B)$ selected by the 5-vectors method after vetoing criteria have been applied.}
    \label{fig:5v_cand_AB}
\end{figure}

Figures \ref{fig:5v_cand_par_space} and \ref{fig:5v_cand_AB} show the outlier distribution in the parameter space of $(f, \dot f)$ and $(A, B)$, respectively. These representations are helpful to visualize the regions defined from equations~(\ref{eq:A_val}) and (\ref{eq:B_val}), which are a rectangle for the physical parameters $(A, B)$ and a parallelogram for $(f, \dot f)$. 

As none of the outliers is loud enough to overcome the threshold, upper limits on the strain of the signal are computed.

\begin{figure}[htb]
    \centering
    \includegraphics[width=\columnwidth]{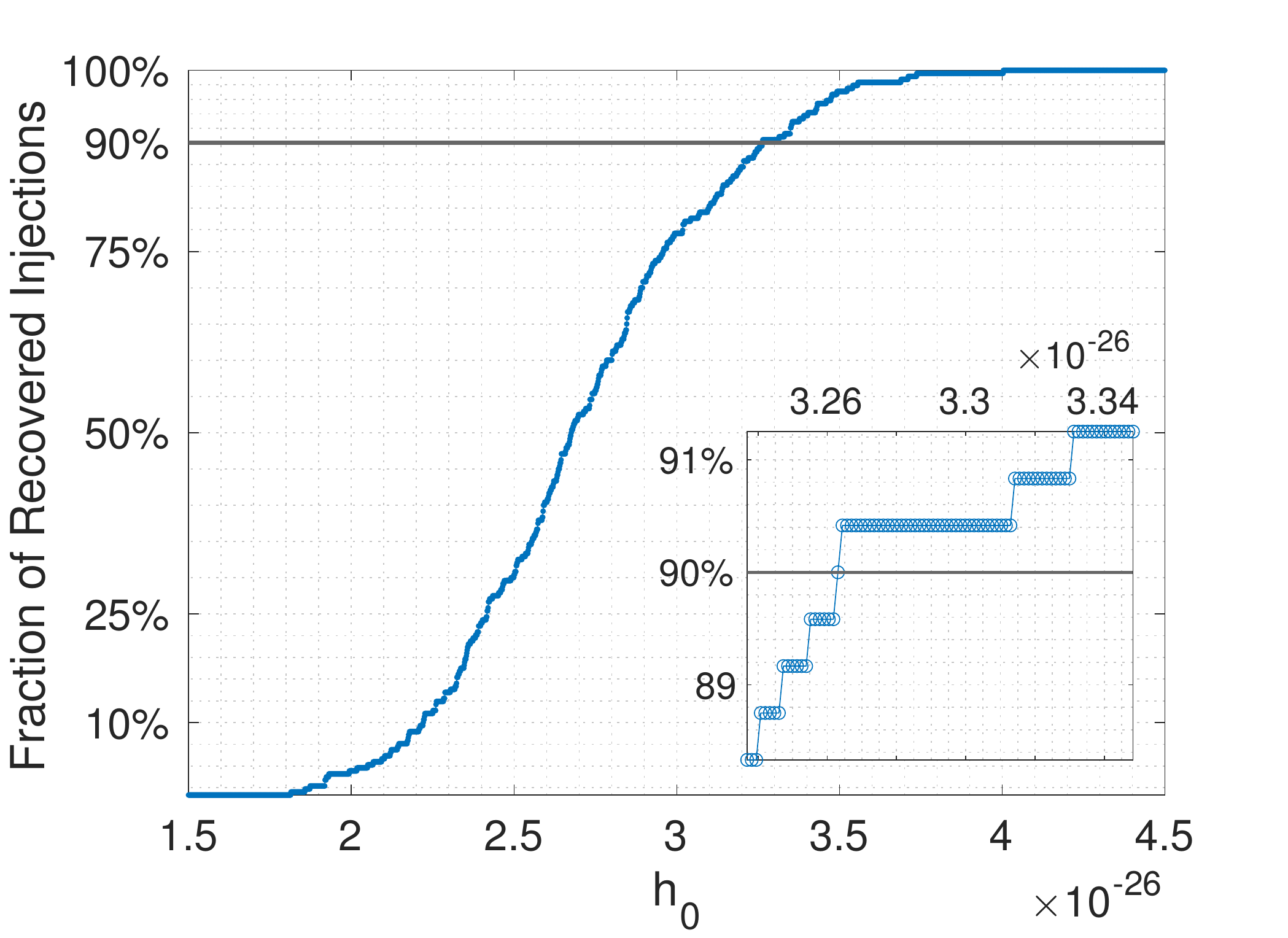}
    \caption{Fraction of recovered injected signals as a function of the strain amplitude $h_0$. The $90\%$ threshold represents the level where the upper limit is identified. On the bottom-right corner, we  show an enlargement of the upper limit region. This example refers to the band 92--92.5~Hz.}
    \label{fig:5v_frac_UL}
\end{figure}

{\em Upper limits.} 
Upper limits at 90$\%$ confidence level are computed every 0.25~Hz by means of software injections based on the procedure already used in standard {\it narrow-band} searches (see, e.g., \citealt{nb_vsr4, nb_O1, nb_O2}). Specifically, for each 0.25~Hz band, sets of 240 signals, with a fixed amplitude and randomly chosen parameters in their allowed range of variation, are injected in O3 data. The signal frequency and spin-down rate are shifted at each glitch epoch in such a way as to simulate  glitches that occurred in the real signal we are searching for. These data are analyzed in exactly the same way as in the real search and the number of detected signals (i.e., those producing a value of the detection statistic above the detection threshold) are counted. By repeating the procedure for different amplitudes, the value corresponding to a detection efficiency of 90$\%$ is obtained through a linear interpolation among the two amplitudes corresponding to a detection probability just below and above 90$\%$, respectively. An example of this procedure for the band 92--92.5~Hz is shown in Figure~\ref{fig:5v_frac_UL}. 

\begin{figure}[htb]
    \centering
    \includegraphics[width=\columnwidth]{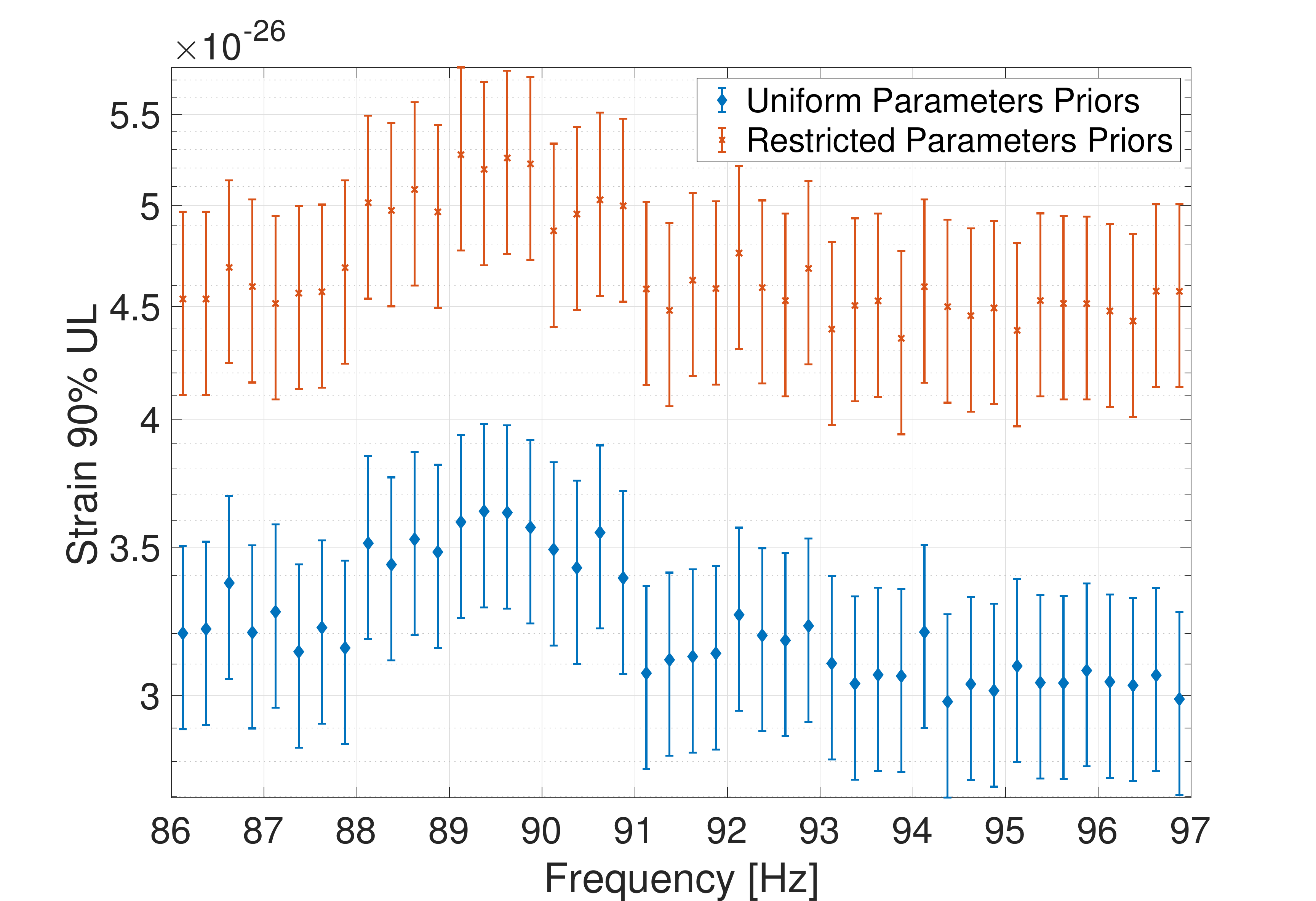}
    \caption{GW strain $90\%$ confidence level upper limits as a function of frequency, obtained using the 5-vectors pipeline. Bottom blue curve corresponds to uniform prior on the polarization angle $\psi$ and the cosine of the star's rotation axis inclination angle $\iota$, while the upper red curve corresponds to ``restricted'' priors based on X-ray observations of the PWN (see text for details).}
    \label{fig:5v_UL}
\end{figure}

The upper limits obtained using this method are shown in Figure~\ref{fig:5v_UL}. The lower blue curve is obtained assuming a uniform prior on the polarization angle $\psi$ and on the cosine of the star's spin axis inclination angle $\iota$. 
On the other hand, the upper red curve assumes a ``restricted'' prior on $\psi$ and $\iota$ based on X-ray observations of the PWN, as discussed in Section~\ref{sect:signal_model}. Specifically, we use for $\psi$ a Gaussian distribution centered at 2.2864 rad with width of 0.0384 rad and a double Gaussian distribution for $\iota$ centered at 1.522 rad and at 1.620 rad, both with standard deviation 0.016 rad. 

The uncertainty on the upper limit is given by the combination (in quadrature) of the statistical error due to the finite number of injections ($\sim 6\%$), the interpolation error in computing the amplitude corresponding to a detection efficiency of 90$\%$ (less than $\sim 1\%$), and the data calibration uncertainty ($\sim 7\%$) (see Section~\ref{sec:gwdata}).

\subsection{Astrophysical constraints}

We now discuss astrophysical consequences of our results. To understand if we are probing a realistic portion of parameter space and constraining theoretical models for r-mode driven spin-down of pulsar \psrname, we compare our upper limits on the amplitude to the spin-down upper limit, i.e., GW amplitude that would be needed to explain the entire measured spin-down of the star
\be
h_{\rm sd}=\frac{1}{d}\sqrt{\frac{10 G}{c^3} I_{zz}\frac{\nu|\dot{\nu}|}{f^2}}
\approx 4.0\times 10^{-26}\left(\frac{90\mbox{ Hz}}{f}\right)\, ,
\label{spindownlimit}
\ee
where $d$ is distance to the source and we use $I_{zz}=10^{45}\mbox{ g cm$^2$}$ and values appropriate for \psrname to obtain the second equality. The spin axis of the star is taken to be in the $z$ direction, and $I_{zz}$ is the moment of inertia. Note that in our case the spin-down limit is not only an upper limit, but it is exactly the predicted amplitude of the signal in the theoretical picture proposed by \citet{Andersson18} to explain the observed inter-glitch braking index of $n=7$. There is, however, an uncertainty associated with equation~(\ref{spindownlimit}), as the relation between $f$, $\nu$, and $I_{zz}$ depends on the unknown EoS of dense matter. 
In order to quantify this uncertainty, we make use of the results of \citet{Idrisy15}, who find that to a good approximation the r-mode (and thus GW) frequency $f$ can be expressed as a function of compactness $\mathcal{C}=GM/c^2 R$, independently of the exact details of the EoS, and neglecting terms of order $(\nu/\nu_K)^2$ (which is a good approximation for the rotation frequency $\nu=62$ Hz of J0537). The amplitude thus depends on only one parameter, which we can take to be the mass of the star $M$ or $I_{zz}$, depending on which is more convenient. 

\begin{figure}[htb]
    \centering
    \includegraphics[width=\columnwidth]{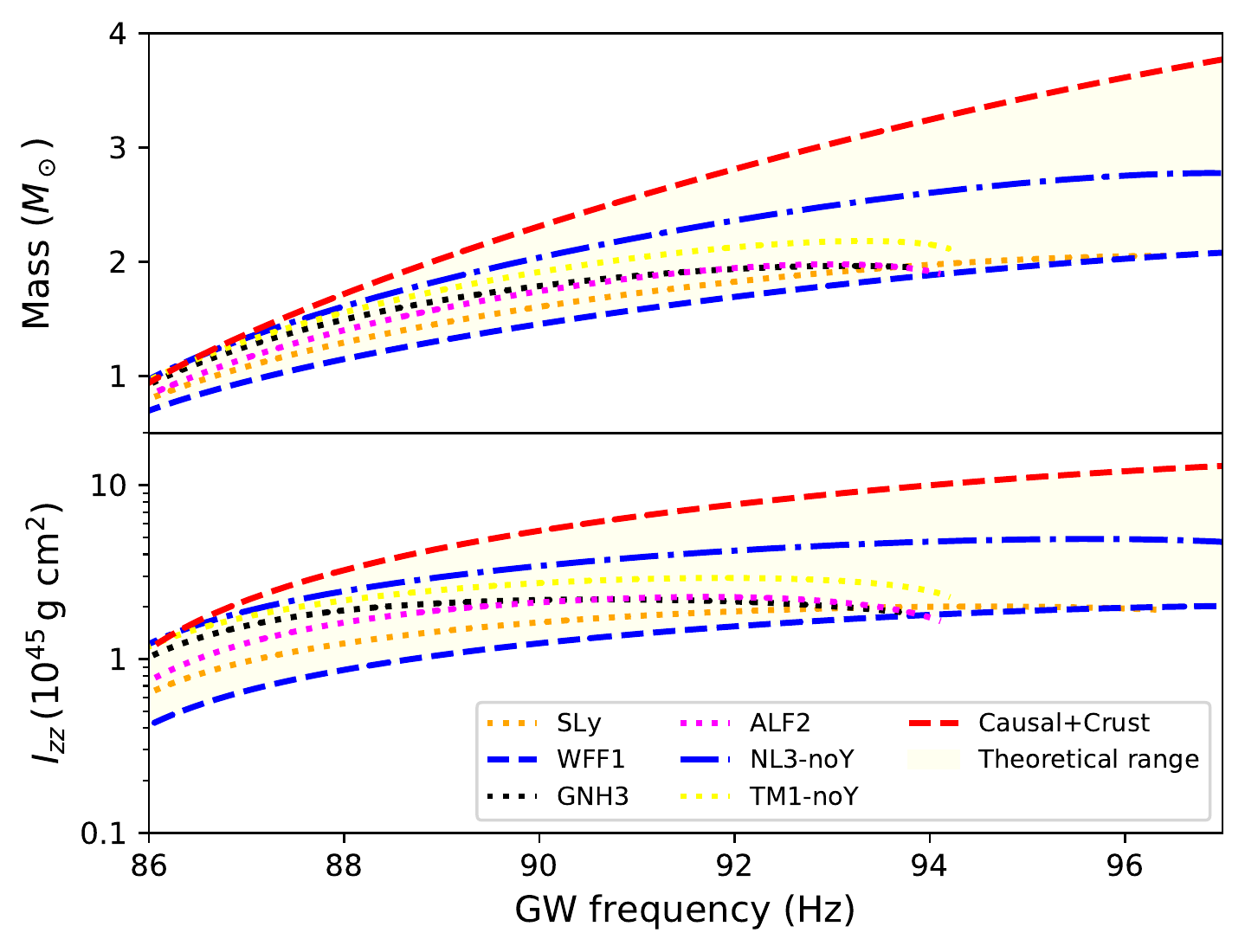}
    \caption{Ranges of theoretically possible mass $M$ and moment of inertia $I_{zz}$ consistent with r-mode emission at the GW frequencies explored in our searches. These ranges reflect our uncertainty in the true equation of state of dense matter, and make no assumption as to whether the observed spindown is driven by r-modes.  The upper limit is set by the stiffest possible EoS from \citet{Haskell18}, which is causally limited in the core and attached to a realistic crustal model, and the lower limit is set by the softest EoS we consider, WFF1, that is still compatible with the observations of a $\approx 2 M_\odot$ neutron star, as described in \citet{Idrisy15}.} 
    \label{fig:Eosrange}
\end{figure}

In Figure \ref{fig:Eosrange}, we show the physically plausible ranges for mass $M$ and moment of inertia $I_{zz}$ for the GW frequencies of our searches. Note that these ranges make no assumption as to whether r-modes actually do account for the observed spin-down; they simply reflect the range of possible values of $M$ and $I_{zz}$ consistent with a given compactness, given our lack of knowledge of the true equation of state of dense matter.  The upper limits on $M$ and $I_{zz}$ are physically well motivated and correspond to a causally limited EoS in the core of the star,  attached to a crustal EoS at lower densities \citep{Haskell18}, i.e., to a hard equation of state. The lower limits are observationally motivated and set by the softest EoSs considered by \citet{Idrisy15}, which are conservatively chosen as to be consistent with the $99.7\%$ confidence limit of the observed mass of $1.97 M_\odot$ of PSR~J1614$-$2230 (\citealt{Demorest10}; note that the mass measurement has since been refined; see \citealt{Arzoumanian18}).

\begin{figure}[htb]
    \centering
    \includegraphics[width=\columnwidth]{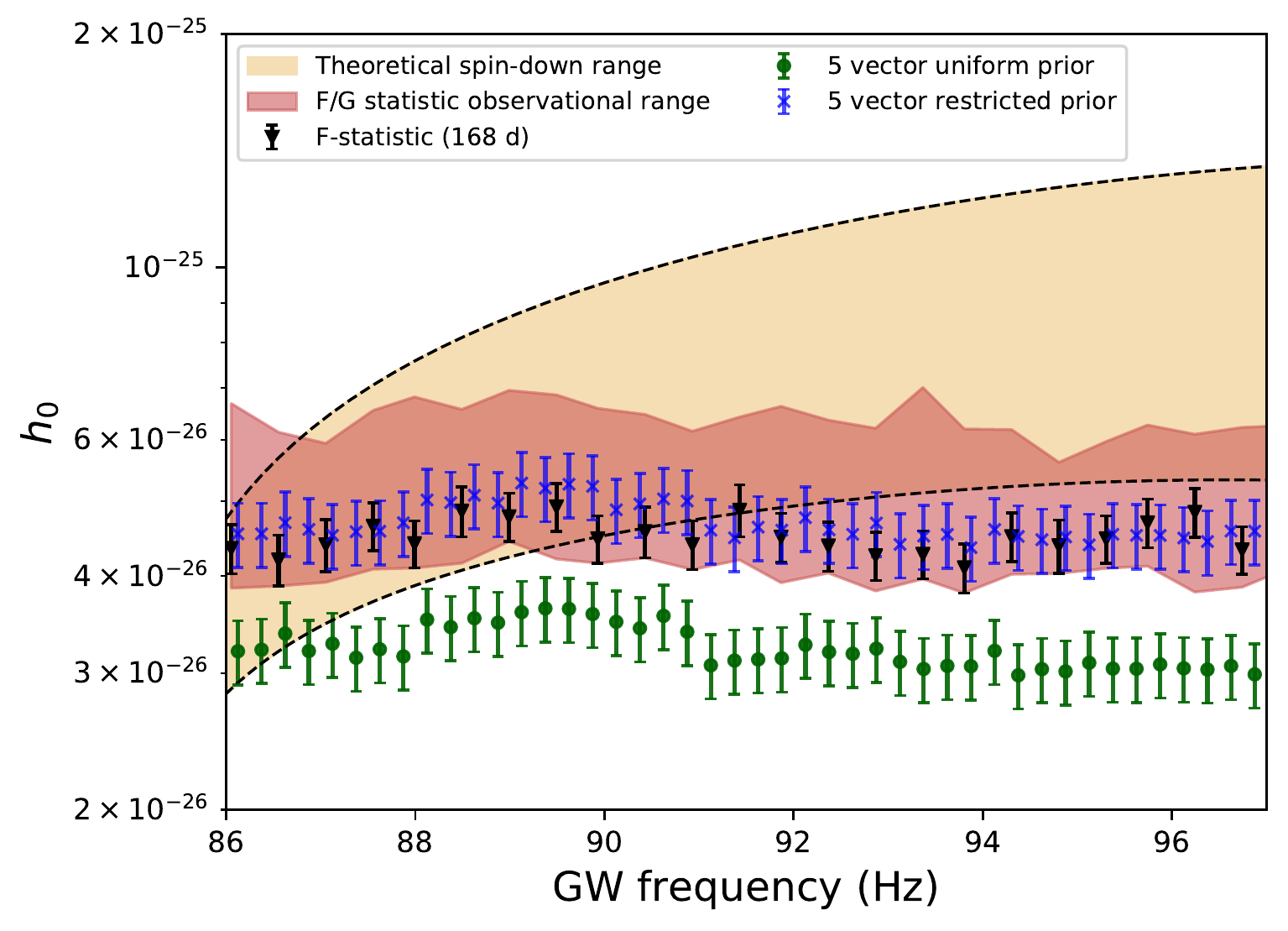}
    \caption{
    Upper limits on GW amplitude $h_0(f)$ obtained from searches using the $\F/\G$ statistic and 5-vector methods. For ease of understanding we plot the results of the $\F$-statistic pipeline only for the longest stretch of data, and indicate the full range of results of the $\F/\G$ statistic pipeline as a shaded band. The dashed lines are set by the stiffest and softest EoSs considered here and enclose a range of theoretical $h_0$  (see text for details).
    }
    \label{fig:SpindownvEoS}
\end{figure}

Having established the physically plausible range for $I_{zz}$, we  calculate the range of $h_{\rm sd}$ from equation~(\ref{spindownlimit}) and compare it to the sensitivity upper limits of our searches.  This is shown in Figure \ref{fig:SpindownvEoS}, where for simplicity we plot only a range between the upper limit set by the stiffest EoS (the causally limited EoS with crust) and the lower limits set by our softest EoS, WFF1. We see that our searches are probing a physically significant portion of parameter space for all frequencies. At frequencies above $f\gtrsim 90$ Hz, our searches are starting to probe below the limits of the r-mode driven spin-down scenario.

\begin{figure*}[htb]
    \centering
    \includegraphics[width=\columnwidth]{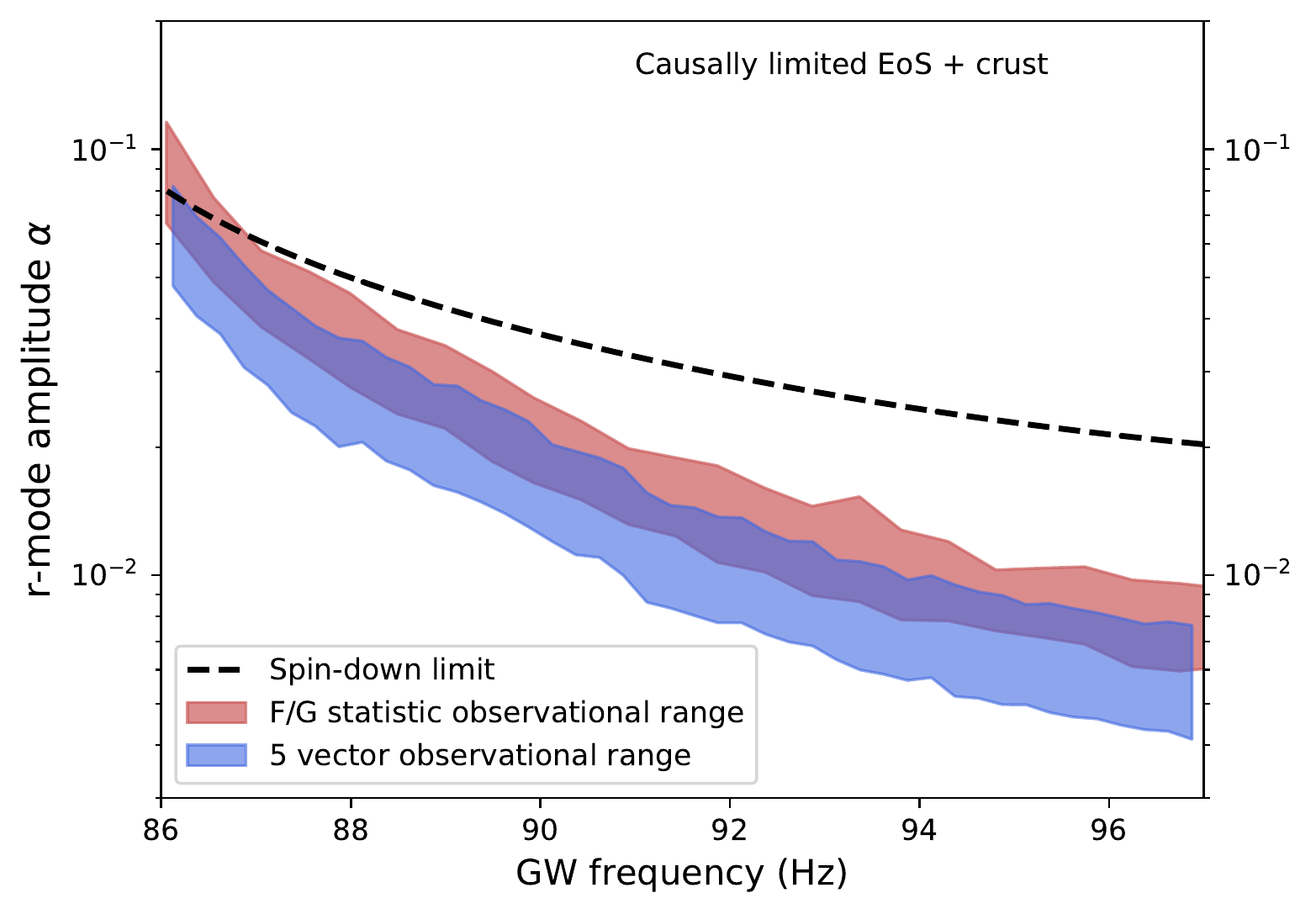}
    \includegraphics[width=\columnwidth]{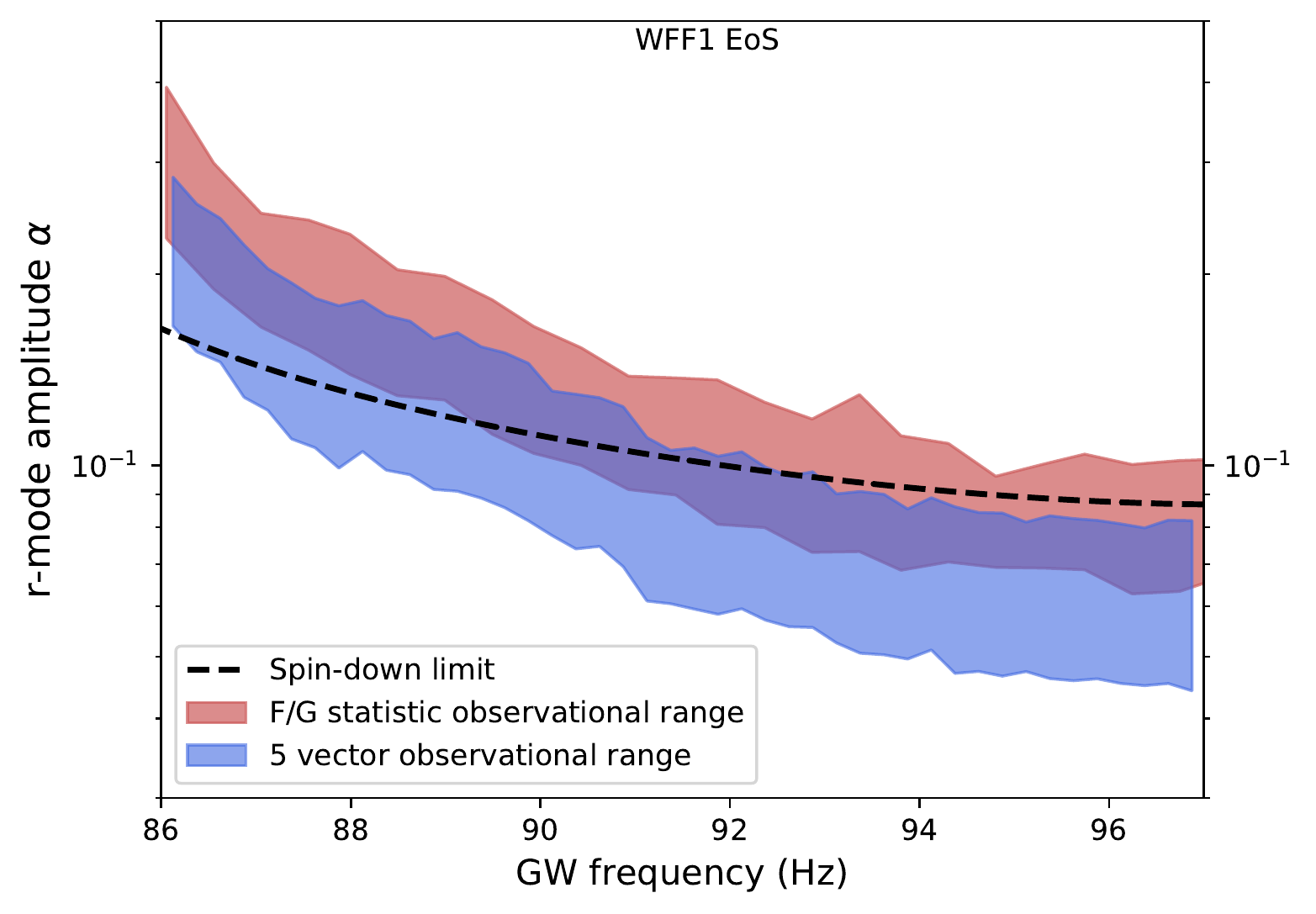}
    \caption{Upper limits on the r-mode amplitude $\alpha$ calculated for two different EoSs. Results for the stiffest EoS we consider, the causally limited EoS with crust, are plotted in the left panel and results for the softest case in our analysis, the WFF1 EoS, are plotted in the right panel. The dotted lines show the amplitude $\alpha$ that would correspond to the spin-down limit.}
    \label{fig:alpalimits}
\end{figure*}

\begin{figure}[htb]
    \centering
    \includegraphics[width=\columnwidth]{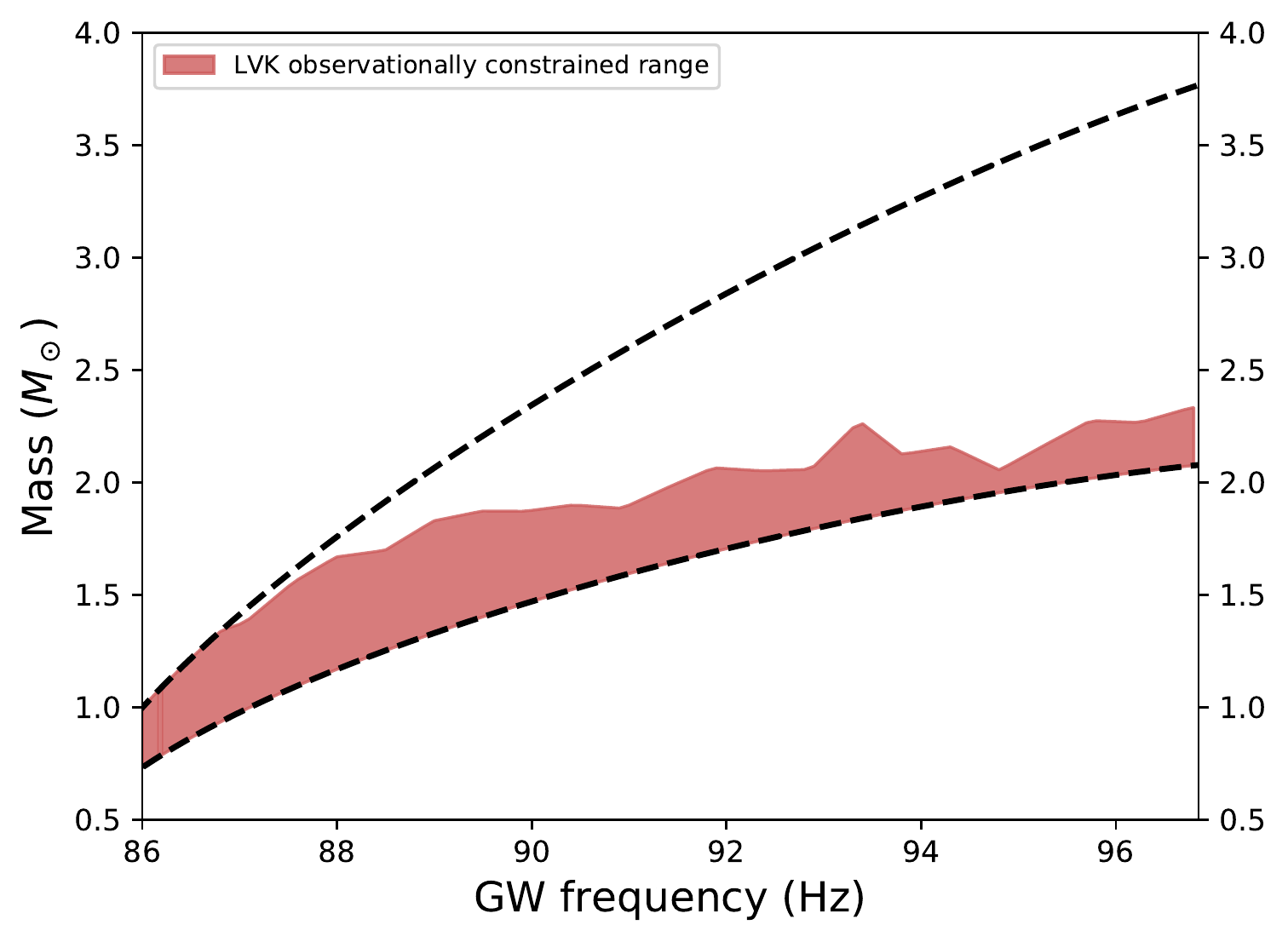}
    \caption{Limits on the mass $M$ of \psrname which is consistent with the assumption that gravitational-wave emission due to r-modes is the dominant mechanism causing the observed spin-down behavior of the pulsar.
    }
    \label{fig:Mass_constraint}
\end{figure}

Another physical quantity of interest is the r-mode amplitude $\alpha$  that leads to a GW signal with an amplitude $h_0$, defined as \citep{Owen:2010}:
\bea
\alpha&=&\sqrt{\frac{5}{8\pi}}\frac{c^5}{G}\frac{h_0}{(2\pi f)^3}\frac{d}{MR^3\tilde{J}} \nonumber\\
&\approx& 0.017\left(\frac{90\mbox{ Hz}}{f}\right)^3\left(\frac{h_0}{10^{-26}}\right),
\eea
where for the second equality we assumed typical values of $M=1.4\,M_\odot$ and $R=12$~km and used the distance to \psrname. $\tilde{J}$ is the dimensionless canonical angular momentum of the r-mode, which depends only weakly on the EoS, and we fix to $\tilde{J}=0.0164$, the value calculated by \cite{Owen1998} for an $n=1$ polytrope. We can use the fit of \cite{Idrisy15} to express the frequency $f$ in terms of compactness and reduce to a single free parameter that encodes the EoS dependence. In Figure~\ref{fig:alpalimits}, we plot upper limits on the amplitude $\alpha$ for our two limiting EoSs, the causally limited EoS with crust and the WFF1 EoS. We see that in both cases the sensitivity of the searches is close to or below the spin-down limit, especially for the stiffest EoS, and is starting to constrain the range of values predicted by theoretical models such as those of \cite{Alford14,Andersson18}. 
We may also visualise the constraints on the r-mode driven spin-down scenario in terms of a constraint on the mass of the neutron star, by using the EoS-independent relations between moment of inertia and compactness of \citet{BreuRezzolla}, together with the fits of \cite{Idrisy15}, so that the only free parameter that encodes the EoS dependence in equation~(\ref{spindownlimit}) is the mass of the star. The results are plotted in Figure~\ref{fig:Mass_constraint} where we see that our observational results allow for the r-mode driven spin-down scenario for soft EoSs and lower-mass neutron stars. Note, however, that for our most sensitive search, the 5-vector search with uniform priors, most of the parameter space is actually excluded, and only the softest EoSs and neutron star masses $M\lesssim 0.9 M_\odot$ are consistent with a GW driven spin-down.

It is also useful to compare the upper limits on $\alpha$ presented above with theoretical estimates of the so-called \emph{saturation amplitude}, i.e. the amplitude at which an unstable r-mode stops growing due to non-linear hydrodynamical couplings with other modes.  The value of the saturation amplitude is highly uncertain and could be even orders of magnitude smaller that the upper limits presented above (see, e.g., \citealt{arras_etal_03, brink_etal_04, bondarescu_etal_09}).  In this sense, the results presented here are consistent with such theoretical estimates.

\section{Discussion}

We presented searches for GWs due to r-modes in the X-ray pulsar \psrname, motivated by theoretical models and observational evidence that the spin-down of the pulsar may be driven by GW emission due to such an unstable oscillation. The search is enabled by a timing ephemeris obtained from \textit{NICER} data \citep{Ho20}, which includes the times of three glitches, and allows us to search in the epochs between glitches, making use of pipelines based on the 5-vector method and the time-domain $\F$/ $\G$ statistic method.

No signal is detected, but we set upper limits on the GW amplitude, which improve by a factor of up to 3 on previous upper limits obtained by \citet{Fesik20,fesikpapa20b} who use data from the O1 and O2 runs of the Advanced LIGO network. The improvement is mostly due to the increased sensitivity of the O3 data analyzed in our searches and to the contemporaneous timing ephemeris obtained from \textit{NICER} data.

The upper limits set by our searches probe a significant portion of parameter space for all frequencies and are beginning to be in tension with the lower limits predicted for the strain \emph{if} GW emission due to an r-mode is indeed driving the entire spin-down of \psrname.
In fact, for our most sensitive search, only a neutron star with mass $M\lesssim 0.9 M_\odot$ is still consistent with the star being spun down by GW emission due to r-modes.
In other words, for a range of frequencies $f\gtrsim 90$ Hz, our searches are probing below the spin-down limit and would be sensitive to emission due to r-modes of lower amplitude than that needed to explain the observations of an inter-glitch braking index of $n=7$ or to the presence of an r-mode that is excited by the glitch but subsequently damped. 
We note, in fact, that we have considered the scenario favored by the models of \citet{Andersson18}, in which the r-mode is always on. This is also the most optimistic scenario for detection.  In the case in which the r-mode is damped, the sensitivity of the search will be degraded, as discussed in detail by \citet{Fesik20} who report a degradation by a factor of $\approx 4$ with respect to the case where the mode is always on; see \citet{Santiago12} for discussion of this scenario.

We conclude by pointing out that it is crucial that \psrname continue to be observed by electromagnetic observatories in order to provide the  ephemeris for future, more sensitive, searches which will be able to securely rule out this theoretical scenario or confirm it with a detection. 
Such a detection would be of great theoretical importance, as it would not only be the first direct detection of GW emission due to r-modes \citep{Andersson98} but also confirm the theoretical suggestion that neutron stars are born rapidly rotating and spun down by GW emission to the currently observed periods of the (non-recycled) pulsar population \citep{Lindblom98,Andersson99, Alford14}.

\section*{Acknowledgments}
The authors gratefully acknowledge the support of the United States
National Science Foundation (NSF) for the construction and operation of the
LIGO Laboratory and Advanced LIGO as well as the Science and Technology Facilities Council (STFC) of the
United Kingdom, the Max-Planck-Society (MPS), and the State of
Niedersachsen/Germany for support of the construction of Advanced LIGO 
and construction and operation of the GEO600 detector. 
Additional support for Advanced LIGO was provided by the Australian Research Council.
The authors gratefully acknowledge the Italian Istituto Nazionale di Fisica Nucleare (INFN),  
the French Centre National de la Recherche Scientifique (CNRS) and
the Netherlands Organization for Scientific Research, 
for the construction and operation of the Virgo detector
and the creation and support  of the EGO consortium. 
The authors also gratefully acknowledge research support from these agencies as well as by 
the Council of Scientific and Industrial Research of India, 
the Department of Science and Technology, India,
the Science \& Engineering Research Board (SERB), India,
the Ministry of Human Resource Development, India,
the Spanish Agencia Estatal de Investigaci\'on,
the Vicepresid\`encia i Conselleria d'Innovaci\'o, Recerca i Turisme and the Conselleria d'Educaci\'o i Universitat del Govern de les Illes Balears,
the Conselleria d'Innovaci\'o, Universitats, Ci\`encia i Societat Digital de la Generalitat Valenciana and
the CERCA Programme Generalitat de Catalunya, Spain,
the National Science Centre of Poland and the Foundation for Polish Science (FNP),
the Swiss National Science Foundation (SNSF),
the Russian Foundation for Basic Research, 
the Russian Science Foundation,
the European Commission,
the European Regional Development Funds (ERDF),
the Royal Society, 
the Scottish Funding Council, 
the Scottish Universities Physics Alliance, 
the Hungarian Scientific Research Fund (OTKA),
the French Lyon Institute of Origins (LIO),
the Belgian Fonds de la Recherche Scientifique (FRS-FNRS), 
Actions de Recherche Concertées (ARC) and
Fonds Wetenschappelijk Onderzoek – Vlaanderen (FWO), Belgium,
the Paris \^{I}le-de-France Region, 
the National Research, Development and Innovation Office Hungary (NKFIH), 
the National Research Foundation of Korea,
the Natural Science and Engineering Research Council Canada,
Canadian Foundation for Innovation (CFI),
the Brazilian Ministry of Science, Technology, and Innovations,
the International Center for Theoretical Physics South American Institute for Fundamental Research (ICTP-SAIFR), 
the Research Grants Council of Hong Kong,
the National Natural Science Foundation of China (NSFC),
the Leverhulme Trust, 
the Research Corporation, 
the Ministry of Science and Technology (MOST), Taiwan,
the United States Department of Energy,
and
the Kavli Foundation.
The authors gratefully acknowledge the support of the NSF, STFC, INFN and CNRS for provision of computational resources.
This work was supported by MEXT, JSPS Leading-edge Research Infrastructure Program, JSPS Grant-in-Aid for Specially Promoted Research 26000005, JSPS Grant-in-Aid for Scientific Research on Innovative Areas 2905: JP17H06358, JP17H06361 and JP17H06364, JSPS Core-to-Core Program A. Advanced Research Networks, JSPS Grant-in-Aid for Scientific Research (S) 17H06133, the joint research program of the Institute for Cosmic Ray Research, University of Tokyo, National Research Foundation (NRF) and Computing Infrastructure Project of KISTI-GSDC in Korea, Academia Sinica (AS), AS Grid Center (ASGC) and the Ministry of Science and Technology (MoST) in Taiwan under grants including AS-CDA-105-M06, Advanced Technology Center (ATC) of NAOJ, and Mechanical Engineering Center of KEK. 
We would like to thank all of the essential workers who put their health at risk during the COVID-19 pandemic, without whom we would not have been able to complete this work.

D.An.\ acknowledges support from an EPSRC fellowship (EP/T017325/1) and NCN grant 2015/18/E/ST9/00577. T.En.\ is supported by JSPS/MEXT KAKENHI grant numbers 15H00845, 17K18776, and 18H04584. C.M.E.\ acknowledges support from ANID FONDECYT/Regular 1171421.  W.C.G.H.\ acknowledges support through grant 80NSSC21K0091 from NASA.

\bibliographystyle{aasjournal}
\bibliography{references} 
\end{document}